\documentclass{aa}   % for short lists of affiliations 
\usepackage{color}
\definecolor{black}{rgb}{0,0,0}
\definecolor{red}{rgb}{1,0,0}
\definecolor{darkblue}{rgb}{0,0,0.7}
\definecolor{blue}{rgb}{0,0,1} 
\definecolor{green}{rgb}{0,0.5,0} 
\definecolor{orange}{rgb}{0.8,0.6,0} 
\definecolor{purple}{rgb}{1,0,1}

\usepackage{orcidlink}
\usepackage{longtable,array}
\usepackage{graphicx}
\usepackage{txfonts}
\usepackage[utf8]{inputenc} % to have accents
\begin{document} 

   \title{Rings around irregular bodies}
      \subtitle{I. Structure of the resonance mesh, applications to Chariklo, Haumea and Quaoar}

\author{
B. Sicardy$^{\orcidlink{0000-0003-1995-0842}}$\inst{1}
\and
H. Salo$^{\orcidlink{0000-0002-4400-042X}}$\inst{2}
\and
M. El Moutamid$^{\orcidlink{0000-0002-4416-8011}}$\inst{3}   
\and
S. Renner$^{\orcidlink{0000-0002-8598-9799}}$\inst{1}
\and
D. Souami$^{\orcidlink{0000-0003-4058-0815}}$\inst{4,5}
}

\institute{
LTE, Observatoire de Paris, Universit\'e PSL, Sorbonne Universit\'e, Universit\'e de Lille, LNE, CNRS 61 Avenue de l'Observatoire, 75014 Paris, France\\
\email{Bruno.Sicardy@obspm.fr}
\and
Space Physics and Astronomy Research unit, University of Oulu, FI-90014 Oulu, Finland
\and
Southwest Research Institute, 
1301 Walnut St, Suite 400, Boulder, CO 80302, 
Boulder, CO 80301, USA
\and
LIRA, CNRS UMR8254, Observatoire de Paris, Universit\'e PSL, Sorbonne Universit\'e, Universit\'e Paris Cit\'e, CY Cergy Paris Universit\'e, Meudon, 92190, France
\and
naXys, Department of Mathematics, University of Namur, Rue de Bruxelles 61, Namur 5000, Belgium
}
 
\date{Received 23 August 2025 / accepted 25 September 2025}

  \abstract
   {Three ring systems have been discovered to date around small irregular objects of the solar system (Chariklo, Haumea and Quaoar). For the three bodies, material is observed near the second-order 1/3 Spin-Orbit Resonance (SOR) with the central object, and in the case of Quaoar, a ring is also observed near the second-order resonance 5/7 SOR. }
   {This suggests that second-order SORs may play a central role in ring confinement. This paper aims at better understanding this role from a theoretical point of view. It also provides a basis to better interpret the results obtained from $N$-body simulations and presented in a companion paper.}
   {A Hamiltonian approach yields the topological structure of phase portraits for SORs of orders from one to five. Two cases of non-axisymmetric potentials are examined: a triaxial ellipsoid characterized by an elongation parameter $C_{22}$ and a body with mass anomaly $\mu$, a dimensionless parameter that measures the dipole component of the body's gravitational field.}
   {The estimated triaxial shape of Chariklo shows that its corotation points are marginally unstable, those of Haumea are largely unstable, while those of Quaoar are safely stable.
   The topologies of the phase portraits show that only first- (aka Lindblad) and second-order SORs can significantly perturb a dissipative collisional ring. 
   We calculate the widths, the maximum eccentricities and excitation time scales associated with first- and second-order SORs, as a function of $C_{22}$ and $\mu$.
   Applications to Chariklo, Haumea and Quaoar using $\mu \lesssim 0.001$ show that the first- and second-order SORs caused by their triaxial shapes excite large ($\gtrsim 0.1$) orbital eccentricities on the particles, making the regions inside the 1/2 SOR inhospitable for rings.
   Conversely, the 1/3 and 5/7 SORs caused by mass anomalies excite moderate eccentricities ($\lesssim$0.01), and are thus a more favorable place for the presence of a ring.}
   {}
   

   \maketitle

\nolinenumbers	% Bruno: si on veut supprimer les num\'eros de lignes

\section{Introduction}
\label{sec_intro}

In the last decade, three dense ring systems have been discovered around small bodies of the solar system. Currently, two rings have been observed around the Centaur object Chariklo \citep{braga2014,sicardy2018}, one ring is known around the dwarf planet Haumea \citep{ortiz2017} and two rings have been detected around the trans-Neptunian Object Quaoar \citep{morgado2023,pereira2023}. Meanwhile, dense and transient material that could be a ring in formation has been detected around the Centaur Chiron \citep{ortiz2023}.

The above mentioned rings differ by a factor of five in terms of orbital radii and heliocentric distances, see the reviews by \cite{sicardy2018} and \cite{sicardy2024b}. Moreover, in the case of Quaoar, the rings are well beyond the classical Roche limit, which challenges the very concept of Roche's zone. Another peculiarity of Quaoar's main ring is that its optical depth significantly varies in longitude, recalling Neptune's ring arc system \citep{depater2018}.

Meanwhile, these rings share common properties. 
They are all dense, in the sense that their optical depths range from about 1\% to more than unity, implying that the particles suffer a few to tens of collisions per revolution. Thus, they must be considered as collisional disks, as opposed to tenuous dusty rings where particles move essentially independently of one another. 
Another common property of these rings is that they are strongly confined over radial distances 
of some kilometers to a few tens of kilometers, calling for an active confining mechanism.
Finally, all these rings orbit close to a second-order resonance with the central body.
More precisely, Chariklo's, Haumea's and Quaoar's main rings are close to the 1/3 resonance, 
meaning that a ring particle completes one revolution when the body completes three rotations,
while Quaoar's fainter ring orbit close the 5/7 resonance, where particles complete five revolutions
during seven rotations of the body.

In this context, we have investigated the behavior of collisional rings around irregular bodies, 
with applications to Chariklo, Haumea and Quaoar.
Our results are presented in two papers. 
The current paper (``Paper I") mainly deals analytical to semi-analytical calculations, focusing on the dynamical structures 
of resonances of various orders around an irregular body.
The second paper by \cite{salo2025} (``Paper~II" hereafter) presents results obtained with $N$-body  simulations of collisional rings perturbed by resonances, and is the numerical counterpart of this paper.

These two papers are more detailed versions of previous works presented 
by \cite{salo2021}, \cite{sicardy2021}, \cite{salo2024} and \cite{sicardy2024a}, where the dense mesh of resonances around irregular objects of the solar system and the importance of the 1/3 SOR for confining rings were pointed out. 

\section{Resonances around an irregular body}
\label{sec_resonance_nomenclature}

We consider a test particle moving in the equatorial plane of a body of mass $M$. The universal gravitational constant is noted $G$ and time is noted $t$, while ${\bf r}$, $r$ and $L$ denote the position vector, the radial distance to the body center of mass and the true longitude of the particle, respectively.
The Keplerian orbital elements of the particle are denoted $a,e,\lambda,\varpi$
(semi-major axis, orbital eccentricity, mean longitude and longitude of pericenter, respectively), 
while $n$ denotes the mean motion of the particle.

In addition to the spherical potential\footnote{Unless otherwise mentioned, the energies and potentials used in this paper are given per unit mass.} $-GM/r$ created by the body, the axisymmetric terms of the potential (e.g. due to the body's oblateness) force a secular apsidal precession rate $\dot{\varpi}_{\rm sec}$ of the particle. 
Moreover, the non-axisymmetric terms create the Spin-Orbit Resonances (SORs) considered in this paper. They may stem from 
a mass anomaly due to topographic features (mountains, craters, etc.), 
a ``mascon" inside the body,
a triaxial shape, or 
more complex shapes. 

The pattern speed of the potential is equal to the spin rate $\Omega_{\rm B}$ of the body.
The orientation of the mass anomaly (or the major axis of the triaxial body) in inertial space is specified by its mean longitude $\lambda' = \Omega_{\rm B} t$. 
The gravitational potential at ${\bf r}$ is then
\begin{equation}
U({\bf r})= \sum_{m=-\infty}^{+\infty} U_m(r) \cos(m\theta),
\label{eq_U0}
\end{equation}
where $\theta= L-\lambda'$.
The terms $U_m(r)$ depend on the particular problem under consideration. 
The expression of $U_m(r)$ for a mass anomaly is given in Appendix~\ref{app_potential_mass_anomaly}, while its expression for a homogeneous triaxial ellipsoid if provided in Appendix~\ref{app_potential_ellipsoid}.

In Eq.~\ref{eq_U0}, we have chosen to vary $m$ from $-\infty$ to $+\infty$ rather than from 0 to $+\infty$, thus implying that $U_m(r)= U_{-m}(r)$. This choice is arbitrary and is made to align with the symmetry in our resonance labeling, where $m$ can be either positive or negative, see below.

Two types of resonances occur around the body. The corotation resonance is defined by
\begin{equation}
n = \Omega_{\rm B},
\label{eq_corot}
\end{equation}
while the $m/(m-j)$ SORs correspond to
\begin{equation}
j\kappa = m(n-\Omega_{\rm B}),
\label{eq_SOR}
\end{equation}
where $\kappa = n  - \dot{\varpi}$ is the epicyclic frequency of the particle. By convention, the integer $j$ (called the order of the resonance hereafter) is always positive. 
In contrast, $m$ can be positive (resp. negative) corresponding to inner (resp. outer ) resonances that occur inside (resp. outside) the corotation radius. 

In this paper, 1$^{\rm st}$-order resonances ($j=1$) are also referred to as Lindblad resonances.
The nomenclature ``$m/(m-j)$ SOR" stems from the fact that Eq.~\ref{eq_SOR} can be re-written as
\begin{equation}
\frac{n-\dot{\varpi}}{\Omega_{\rm B}-\dot{\varpi}}= 
\frac{m}{m-j} \approx
\frac{n}{\Omega_{\rm B}},
\label{eq_ratio_n_omega_exact}
\end{equation}
where the approximation is valid only if $\dot{\varpi} \ll n, \Omega_{\rm B}$, 
which is usually the case in planetary problems.
In galactic dynamics, this is not true anymore, so the notation ``$m/(m-j)$ resonance"
becomes meaningless. Another notation -- not to be confounded with the one adopted here -- 
is then used, for instance ``$m:1$ Lindblad resonance" for the case $j=1$ (see e.g. \citealt{pfenniger1984}).

The potential $U({\bf r})$ can be expressed in terms of the Keplerian elements of the particle and 
Fourier-expanded under the form
\begin{equation}
\displaystyle
U(a,e,\lambda,\varpi,\lambda') = \sum_{m=-\infty}^{+\infty} \sum_{j=0}^{+\infty} 
\overline{U}_{m,j}(\alpha) e^j \cos (\psi_{m,j}),
\label{eq_pot_expan}
\end{equation}
where 
\begin{equation}
\psi_{m,j}= m\lambda' - (m-j)\lambda - j\varpi
\label{eq_psi_m_j}
\end{equation}
and $\alpha= a/R_{\rm ref}$, where $R_{\rm ref}$ is a reference radius that gives the characteristic size of the object, for instance its radius if it is a sphere. More complex expressions are obtained for a triaxial object, see Eq.~\ref{eq_R_ref}.
In each term of the sum, we have kept only the lowest order term in eccentricity, that is $e^j$.
The terms $\overline{U}_{m,j}(\alpha)$'s are given by
\begin{equation}
\overline{U}_{m,j} (\alpha)= 2F_N [U_m(\alpha)],
\label{eq_Umj_gen}
\end{equation}
where the $F_N$'s are linear operators acting on $U_m(\alpha)$ that contain only multiplicative factors and derivatives with respect to $\alpha$ up to degree $j$. They are labeled by the index $N$, according to the nomenclature of \cite{murray2000} or \cite{ellis2000}, see \cite{sicardy2020a} for details and Appendix~\ref{app_strength_resonances} for a summary. The operators $F_N$'s are listed in Table~\ref{tab_Fn} for 1$^{\rm st}$- and 2$^{\rm nd}$-order SORs only, because higher order resonances are not expected to have a significant effect on a collisional disk, as will be shown later.
We point out that in the case of a mass anomaly and $|m|=1$, the potential $U_m(\alpha)$
contains an indirect term (Eq.~\ref{eq_U_m_alpha_ma}) that is automatically included to calculate 
$\overline{U}_{m,j} (\alpha)$ from Eq.~\ref{eq_Umj_gen}.

Caution must be taken with the factor two appearing in Eq.~\ref{eq_Umj_gen}. It should be used only if the azimuthal number $m$ appearing in Eq.~\ref{eq_U0} varies from $-\infty$ to $+\infty$, as we do here. If $m$ varies 0 to $+\infty$ in Eq.~\ref{eq_U0}, as is the case in the literature for the potential of a satellite, then Eq.~\ref{eq_Umj_gen} must be replaced by $\overline{U}_{m,j} (\alpha)= F_N [U_m(\alpha)]$.

We note that Eq.~\ref{eq_Umj_gen} can be used for any potential with the form of Eq.~\ref{eq_U0}. 
It has the advantage of encapsulating in a single formula inner and outer resonances including possible indirect terms, or even retrograde resonances if $n/\Omega_{\rm B} < 0$. 

\section{Jacobi constant and phase portraits}
\label{sec_J_and_phase_portraits}

Appendix~\ref{app_hamiltonians} is a summary of how the classical Hamiltonians corresponding to SORs of any order $j$ are obtained. It provides all the expressions that will be necessary from this section through Section~\ref{sec_J_e}.
We first note that the phase portraits of the resonances are parameterized by the Jacobi constant $J$. We will use hereafter two equivalent forms of $J$. One is the dimensionless form
\begin{equation}
\Delta J = \frac{1}{2}  \left[ \frac{\Delta a}{a_0} + \left( \frac{m-j}{j} \right) e^2\right],
\label{eq_definition_Delta_J}
\end{equation}
where $\Delta a= a-a_0$ and $a_0$ is the semi-major axis at ``exact" resonance, 
i.e. where the condition~\ref{eq_SOR} is met.
Thus, the condition $\Delta J=0$ may be loosely viewed as a definition of the center of the resonance.
The Jacobi constant can also be expressed through the quantity
\begin{equation}
\overline{a} = a_0(1 + 2\Delta J) = a + a_0 \left( \frac{m-j}{j}\right) e^2,
\label{eq_definition_Delta_J_alt}
\end{equation}
which has the dimension of a length. It will hereafter be referred to as the ``modified semi-major axis".
It corresponds to the semi-major axis of the circular orbit of a particle that has the constant of motion $\Delta J$.
The main advantage of using $\overline{a}$ over $a$ is that it is constant of motion time for a particle in the $m/(m-j)$ resonance, once high frequency terms have been averaged out. Meanwhile, since $e$ is generally small, it also gives a good assessment of the particle's semi-major axis. 

\begin{figure*}[h!]
\centerline{\includegraphics[totalheight=40mm,trim=0 0 0 0]{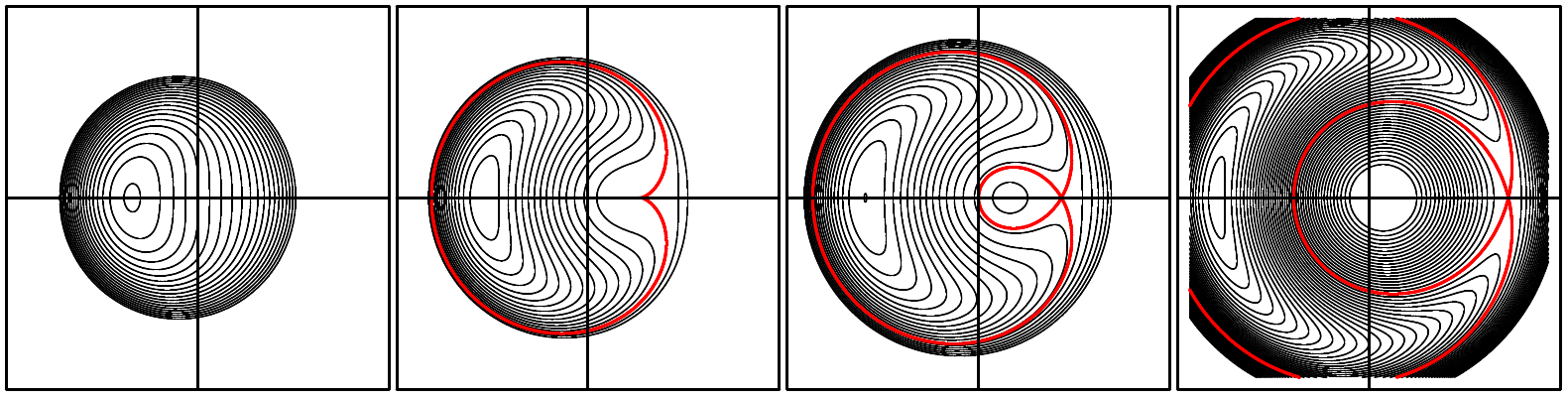}}
\centerline{\includegraphics[totalheight=40mm,trim=0 0 0 0]{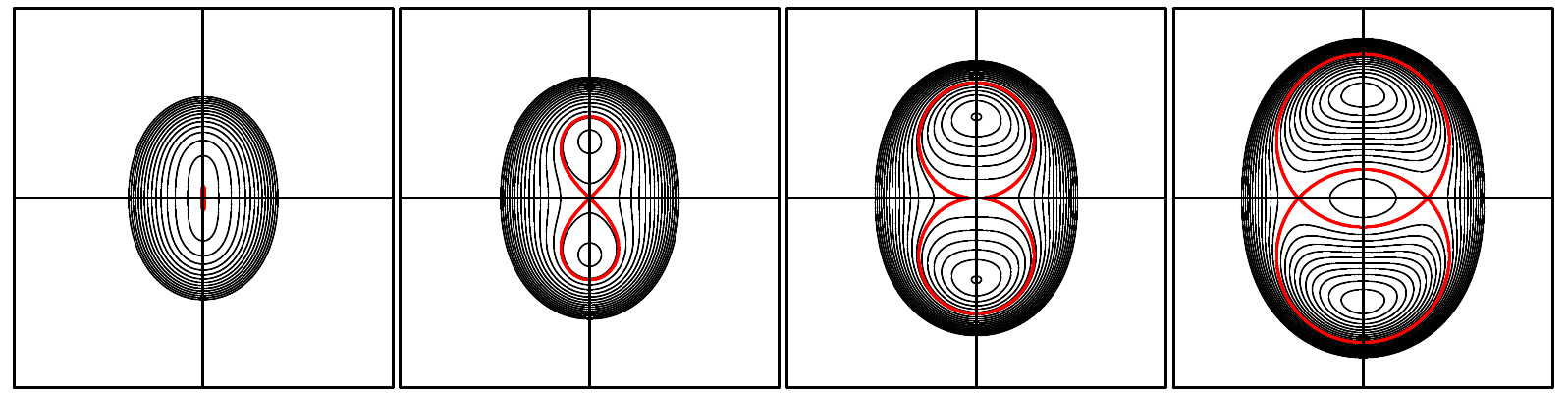}}
\centerline{\includegraphics[totalheight=40mm,trim=0 0 0 0]{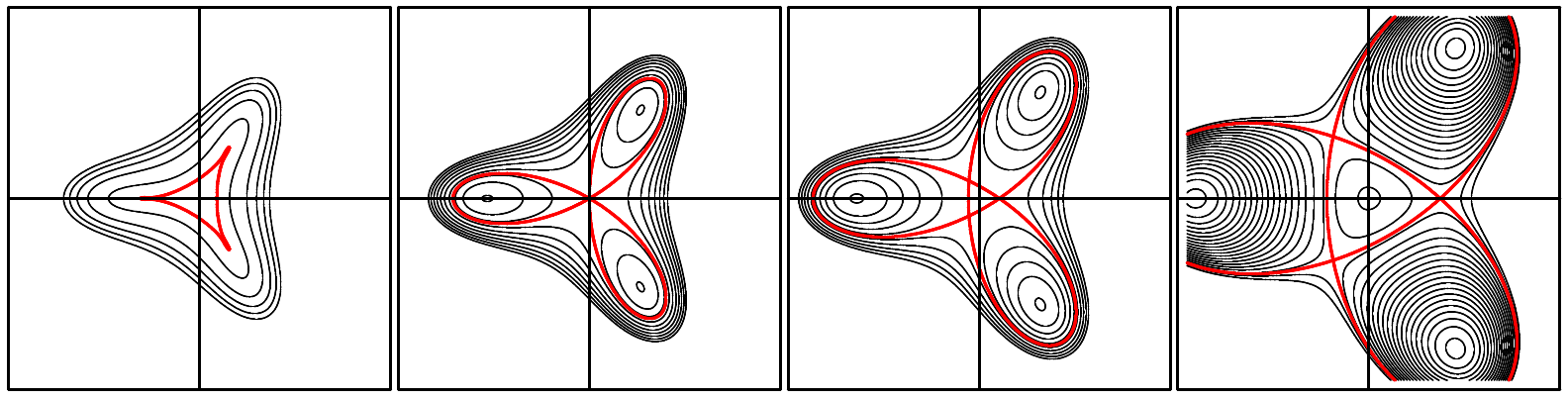}}
\centerline{\includegraphics[totalheight=40mm,trim=0 0 0 0]{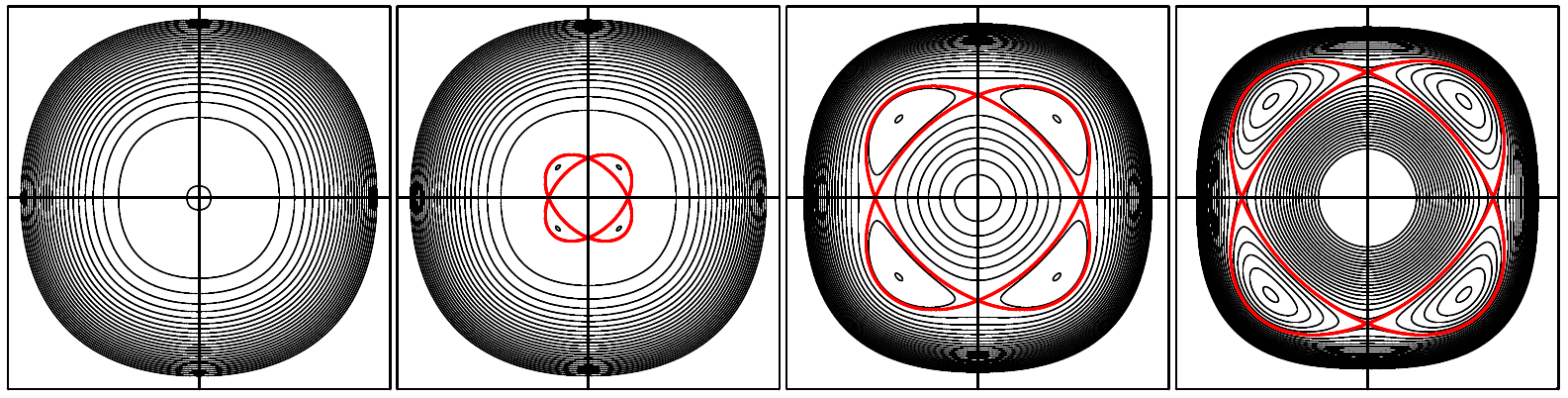}}
\centerline{\includegraphics[totalheight=40mm,trim=0 0 0 0]{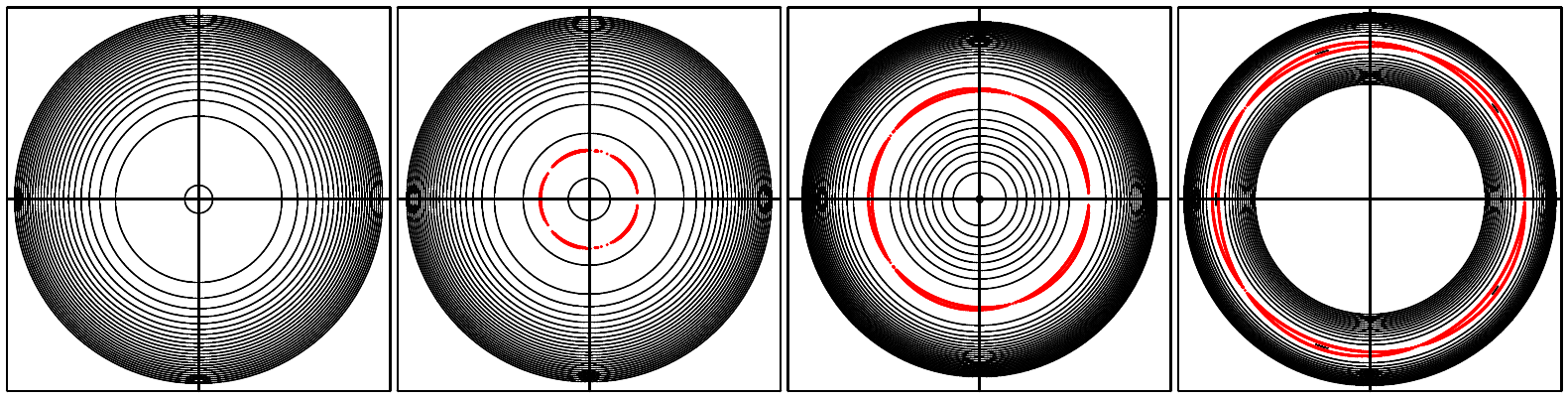}}
\caption{
Representative phase portraits of resonances with orders $j$=1, 2, 3, 4 and 5, from top to bottom, respectively.
Each phase portrait shows level curves of the Hamiltonians ${\cal H} (X,Y)$ given in Eq.~\ref{eq_H_X_Y}, where the mixed variables
$X$ and $Y$ (Eq.~\ref{eq_def_X_Y}) define the eccentricity vector ${\bf e}$ (Eq.~\ref{eq_ecc_vector}).
The fixed elliptic points away from the origins correspond to maxima of ${\cal H} (X,Y)$.
For each resonance, four representative values of $\Delta J$ decreasing from left to right have been considered to illustrate the varying topologies of the phase portraits.
The homoclinic trajectories are drawn in red.
In all the plots, the value of the parameter $\epsilon$ appearing in Eq.~\ref{eq_H_X_Y}
is taken as negative, as is the case for outer resonances.
The topology of resonances with orders $j > 5$ are similar to the case $j$=5, 
except that there are $j$ islands instead of five, with widths that decrease as $j$ increases.
}
\label{fig_maps_hamiltonians}
\end{figure*}

For a given value of $\Delta J$ (or $\overline{a}$), a test particle evolves in a determined phase portrait, as illustrated in Fig.~\ref{fig_maps_hamiltonians}. 
We distinguish two kinds of periodic orbits. The first-kind orbits correspond to fixed points at the origin of the phase portrait ($e=0$), while the second-kind orbits correspond to fixed points with $e \neq 0$.

In a collisional ring perturbed by a SOR, two opposite trends are at work.
Collisions tend to damp eccentricities, and thus push the particles towards the origin of the phase portrait, which corresponds to a downward motion in the $(\overline{a},e)$ space.
Conversely, the SORs tends to take the particles towards second-kind orbits while maintaining $\overline{a}$ constant, thus corresponding to an upward vertical motion in the $(\overline{a},e)$ space. 

We examine in the next two sections the dynamical stability of first-kind orbits and the locations of the second-kind orbits in the phase portraits.

\section{Stability of first-kind orbits}
\label{sec_stability_1st_kind}

To within a constant factor that is ignored, the Hamiltonian describing the $m/(m-j)$ resonance (Eq.~\ref{eq_H_Theta_phi}) can be re-written as
\begin{equation}
{\cal H} =
3 \Delta J \left( \frac{m-j}{2j} \right) e^2 - \frac{3}{2} \left( \frac{m-j}{2j} \right)^2 e^4 + \epsilon e^j \cos(j\phi),
\label{eq_hamil_expanded_in_e}
\end{equation}
where the resonant angle $\phi$ 
is\footnote{To alleviate the notation, and because $\phi$ is used many times in this paper, we omit the indices $m$ and $j$ that should be attached to it. This should be remembered in all the expressions where $\phi$ appears.}
\begin{equation}
\phi = \frac{\psi_{m,j}}{j}= \frac{m\lambda' - (m-j)\lambda - j\varpi}{j}.
\label{eq_phi}
\end{equation}
The parameter $\epsilon$ quantifies the strength of the resonance. 
It depends on $m$ and $j$ and is defined by
\begin{equation}
\epsilon = \frac{\overline{U}_{m,j}(\alpha)}{a_0^2 n_0^2}.
\label{eq_epsilon}
\end{equation}
see Appendices~\ref{app_strength_resonances} and \ref{app_hamiltonians} for details.
 
\textbf{\textit{Case $\boldsymbol{j=1}$.}}
The Hamiltonian can be expressed in terms of the mixed variables  $X= e\cos(\phi)$ and $Y= e\sin(\phi)$. At the origin of the phase portrait ($X=Y=0$), Eq.~\ref{eq_motion_X_Y} yields $\dot{X} = 0$ and $\dot{Y}= \epsilon \neq 0$. Thus, for 1$^{\rm st}$-order resonances the origin of the phase portrait is never a fixed point. A particle placed on a circular orbit will always sees its orbital eccentricity initially increase, see Fig.~\ref{fig_maps_hamiltonians}.

For higher order resonances ($j \geq 2$), ${\cal H}$ is of order of at least two in eccentricity, 
i.e. contains a homogeneous polynomial $P(X,Y)$ of at least degree two (Eq.~\ref{eq_H_X_Y}).
Consequently, the origin of the phase portrait is always a fixed point.
However, the nature of the this point (elliptic vs. hyperbolic) depends on $j$.

\textbf{\textit{Case $\boldsymbol{j=2}$.}}
To lowest order in eccentricity, and from Eq.~\ref{eq_hamil_expanded_in_e}, we have near the origin
$$
{\cal H} \approx
\left[ \frac{3}{4} (m-2) \Delta J  + \epsilon \cos(2\phi) \right] e^2.
$$
Thus, in the finite interval of Jacobi constant
\begin{equation}
-\frac{4}{3} \left| \frac{\epsilon}{m-2} \right| < \Delta J < +\frac{4}{3} \left| \frac{\epsilon}{m-2}\right|,
\label{eq_interval_DJ}
\end{equation}
the sign of ${\cal H}$ changes along two directions as $\phi$ varies from 0 to $2\pi$.
The origin is then a fixed hyperbolic (unstable) point with two homoclinic trajectories along the directions
defined by  $3(m-2) \Delta J/4  + \epsilon \cos(2\phi) = 0$.
A particle launched on a circular orbit with those values of $\Delta J$
will have its orbital eccentricity increased in a first phase (Fig.~\ref{fig_maps_hamiltonians}).

Outside the interval given above, the origin of the phase portrait is a fixed elliptic (stable) point.
A particle launched on a circular orbit will remain on this circular orbit.

\textbf{\textit{Case $\boldsymbol{j=3}$.}}
To lowest orders in eccentricity, we have again near the origin
$$
{\cal H} \approx
\frac{1}{2} (m-3) \Delta J  e^2 + \epsilon e^3 \cos(3\phi).
$$
If $\Delta J \neq 0$, the second-order term dominates the expression of ${\cal H}$,
so that the origin is an elliptic point.
If $\Delta J = 0$, ${\cal H}$ is dominated by the third-order term. The Hamiltonian changes its sign along the three homoclinic directions defined by $\cos(3\phi)=0$ (Fig.~\ref{fig_maps_hamiltonians}). However, and contrarily to the 2$^{\rm nd}$-order resonances case, this happens only for an isolated value of $\Delta J$. 

\textbf{\textit{Case $\boldsymbol{j=4}$.}}
To lowest orders in eccentricity, we have near the origin
$$
{\cal H} \approx
\frac{3}{8} (m-3) \Delta J  e^2 + \left[ \epsilon  \cos(4\phi) - \frac{3}{128}(m-4)^2 \right] e^4.
$$
If $\Delta J \neq 0$ the origin is an elliptic point.
For $\Delta J = 0$, the origin remains an elliptic point as long as $\epsilon$ remains in the interval
\begin{equation}
-\frac{3}{128}(m-4)^2 < \epsilon < +\frac{3}{128}(m-4)^2.
\label{eq_range_epsilon_stability_origin_j_4}
\end{equation} 
In the opposite case, the origin is an hyperbolic point with four homoclinic directions.
However, this requires $|\epsilon|$ to be quite large, a situation usually not encountered.

\textbf{\textit{Case $\boldsymbol{j \geq 5}$.}} Near the origin, the Hamiltonian is now dominated either by isotropic terms of order two ($\Delta J \neq 0$) or order four ($\Delta J = 0$) in eccentricity, so that the origin is always an elliptic point.

\section{Second-kind orbits}
\label{sec_2nd_kind_orbits}

The second-kind (or resonant) orbits are given by the fixed points of the phase portraits with $e \neq 0$, as shown in Fig.~\ref{fig_maps_hamiltonians}. At these points, $\partial {\cal H}/\partial \phi = \partial {\cal H}/\partial \Theta = 0$. In particular $\partial {\cal H}/\partial \Theta = 0$ yields $\sin (j\phi) = 0$, so that the fixed points lie along the directions defined by
$$
\varphi_k = \frac{k\pi}{j} ~~(k=0,...,2j-1),
$$
in the phase portrait, where $\cos (j\varphi_k) = (-1)^k$.
We define the ``$\varphi_k$-axis" as the line which makes an angle $\varphi_k$ with the $X$-axis,
so that the 0-axis is the $X$-axis, the $\pi/2$-axis is the $Y$-axis, etc.

The equation $\partial {\cal H}/\partial \phi = 0$ then provides the modulus $e$ of eccentricity vector corresponding to the fixed points, i.e.
\begin{equation}
e^2 = \frac{2j}{m-j} \left[ \Delta J + \frac{(-1)^k j^2 \epsilon}{3(m-j)} e^{j-2}  \right] ~~(k= 0,...,2j-1).
\label{eq_e_fixed_gen}
\end{equation}
This equation can be projected onto the $\varphi_k$-axis, yielding
\begin{equation}
E^2 =  \left( \frac{2j}{m-j} \right) \Delta J + (-1)^k \epsilon' E^{j-2} ~~(k= 0,...,2j-1),
\label{eq_e_fixed_E}
\end{equation}
where $E$ is an algebraic (positive or negative) quantity representing the eccentricity.
In order to simplify the expressions obtained hereafter,
we introduce a change of variable which writes
\begin{equation}
\epsilon'= \frac{2 j^3}{3(m-j)^2} \epsilon.
\label{eq_epsilonp}
\end{equation}

For $j=1$, it is sufficient to consider the case $k=0$, corresponding to fixed points along the $X$-axis.
For $j \geq 2$, it is enough to consider the cases $k=0$ and $k=1$, 
as all the remaining cases $k=2,...,2j-1$ are a mere repetition of Eq.~\ref{eq_e_fixed_E},
due to the invariance of the Hamiltonian under rotations of $2\pi/j$ radians.

\begin{figure*}[t!]
\centerline{
\includegraphics[totalheight=53mm,trim=0 0 0 0]{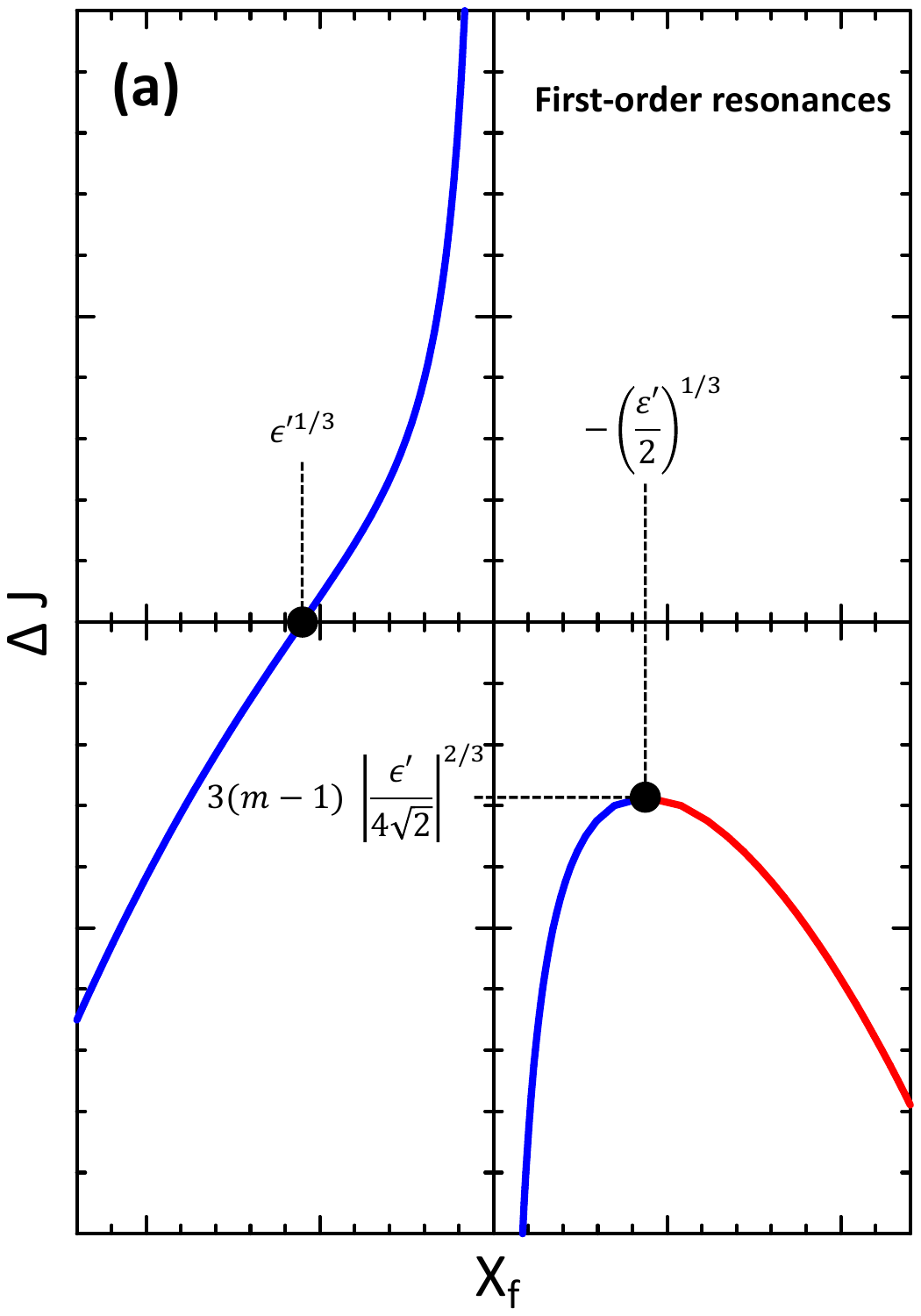}
\includegraphics[totalheight=53mm,trim=0 0 0 0]{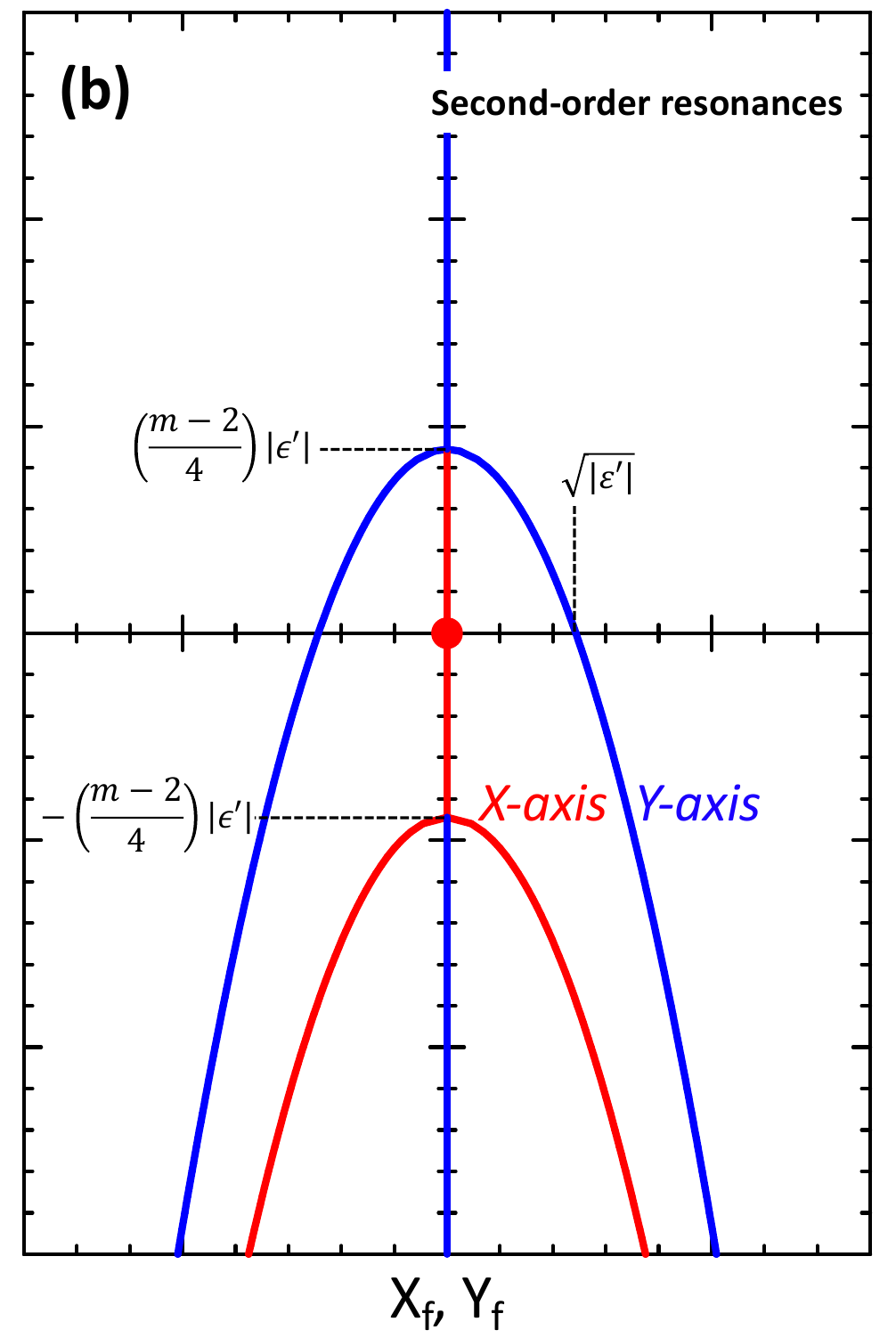}
\includegraphics[totalheight=53mm,trim=0 0 0 0]{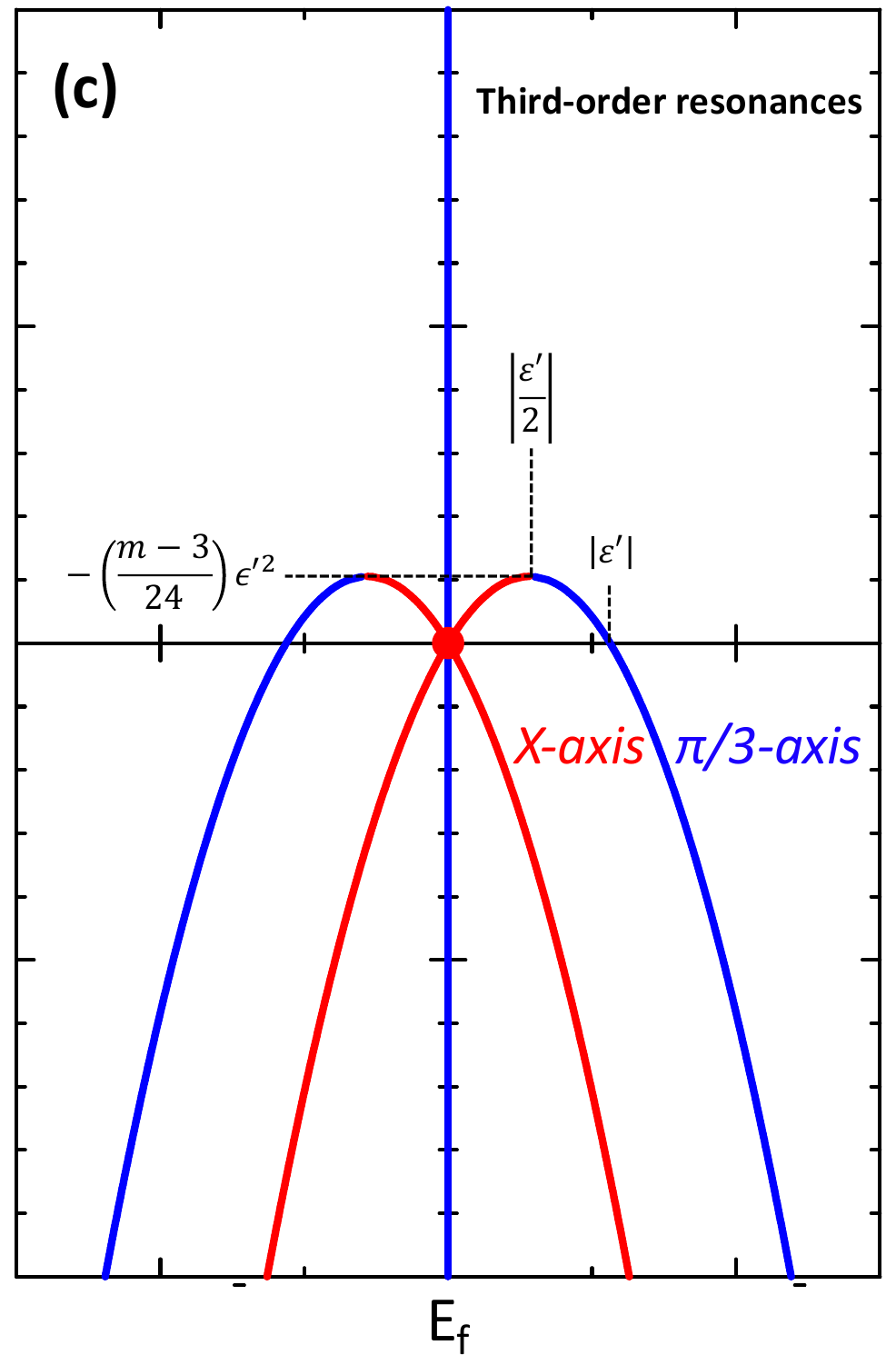}
\includegraphics[totalheight=53mm,trim=0 0 0 0]{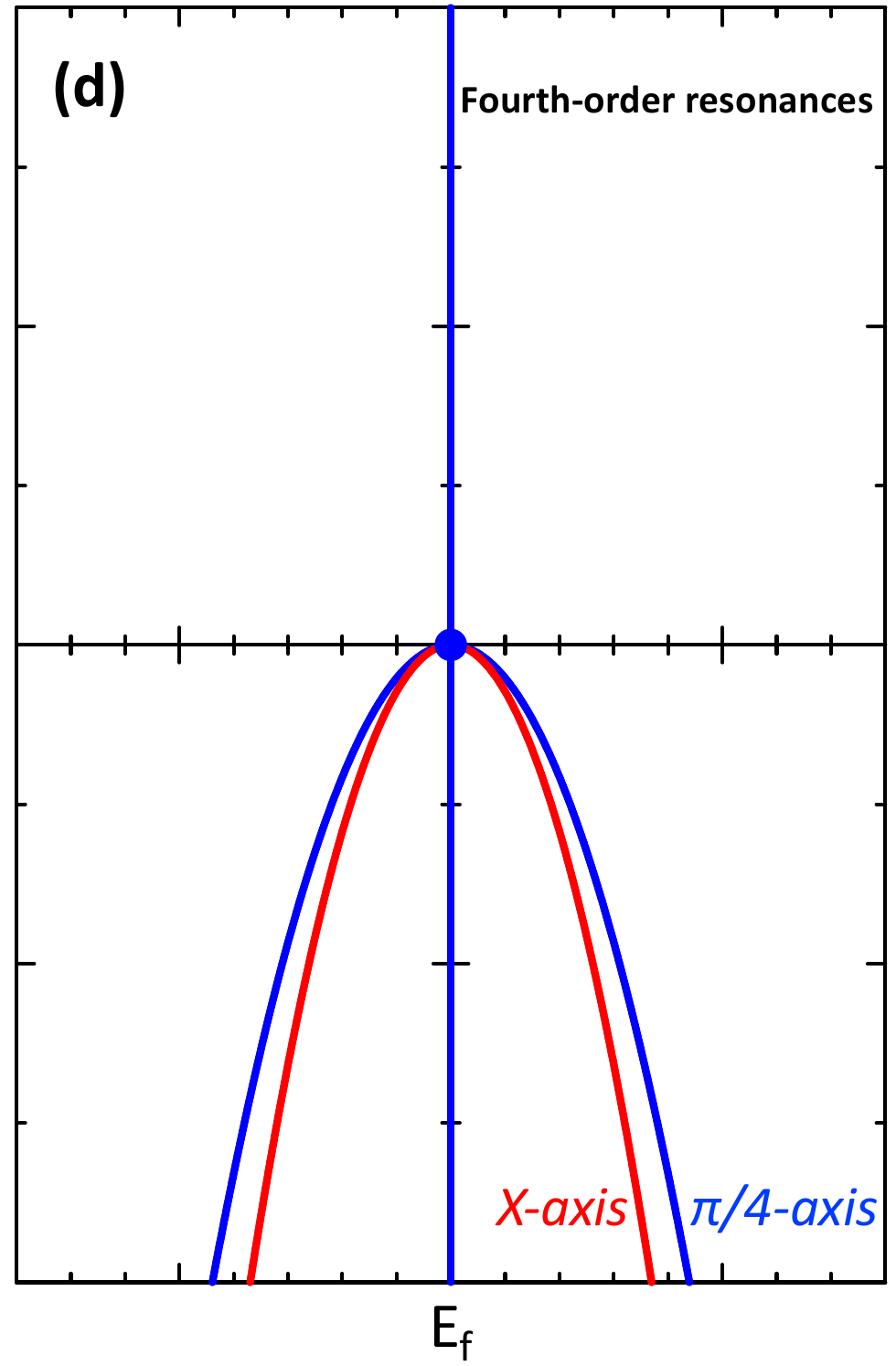}
\includegraphics[totalheight=53mm,trim=0 0 0 0]{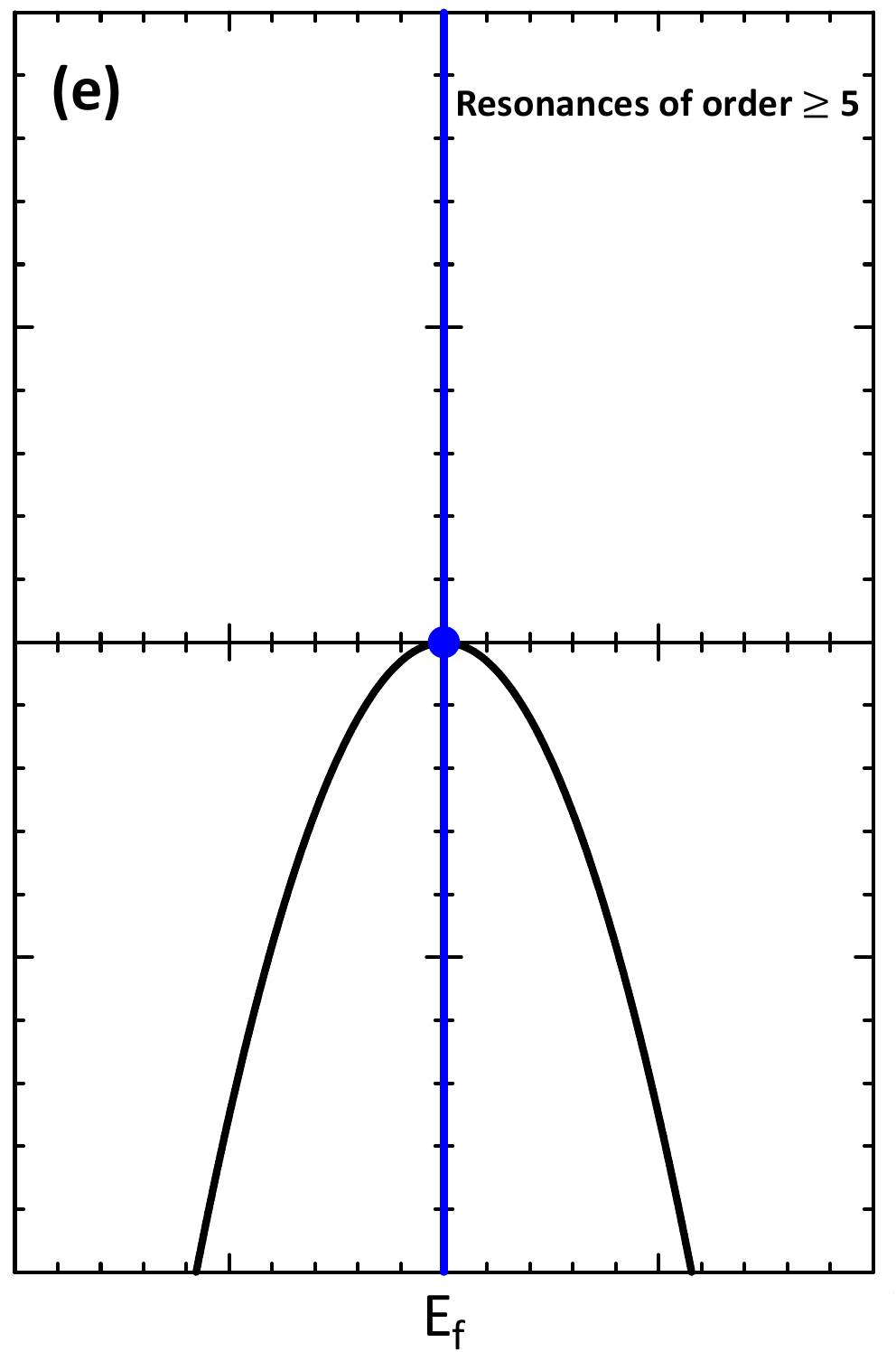}
}
\caption{
Fixed points given by Eq.~\ref{eq_e_fixed_E} as a function of $\Delta J$ for resonances of various orders. They are the solutions of Eqs.~\ref{eq_roots_j_1}, \ref{eq_roots_j_2}, \ref{eq_roots_j_3}, \ref{eq_roots_j_4} and \ref{eq_roots_j_geq_5}.
In all the plots, the values of $\epsilon'$ are taken as negative, as is the case for outer resonances.
The cubic root $\epsilon'^{1/3}$ is then understood as the real root, i.e. ignoring the complex roots.
The blue (resp. red) branches corresponding to stable elliptic (resp. unstable hyperbolic) points.
Similarly, blue (resp. red) dots at the origin indicate a stable (resp. unstable) point.
The units of all the plots are arbitrary.
\textit{Panel~(a)}: first-order resonances.
The positions of two particular points are specified: the solution corresponding to $\Delta J=0$ and 
the pitchfork bifurcation point at the lower right. 
\textit{Panel~(b)}: second-order resonances.
The parabolic branches are the solutions of Eq.~\ref{eq_roots_j_2}. 
The positions of three particular points are specified.
\textit{Panel~(c)}: third-order resonances.
The parabolic branches are the solutions of Eq.~\ref{eq_roots_j_3}. 
The positions of three particular points are specified.
The origin of the phase portrait  ($X_{\rm f} = E_{\rm f}$ =0) is stable everywhere, except for the value $\Delta J = 0$
(red dot), where it is hyperbolic.
\textit{Panel~(d)}: fourth-order resonances.
The parabolic branches are the solutions of Eq.~\ref{eq_roots_j_4}. 
The origin of the phase portrait  ($X_{\rm f} = E_{\rm f}$ =0) is stable everywhere, 
except for large values of $\epsilon$, see Eq.~\ref{eq_range_epsilon_stability_origin_j_4}.
\textit{Panel~(e)}: resonances of orders $j \geq 5$.
In this case, $X_{\rm f} = E_{\rm f}$ (Eq.~\ref{eq_roots_j_geq_5}), 
so that the stability of the points corresponding to each branch cannot be indicated on the plot,
hence the black color used here. 
This plot is now indistinguishable from the unperturbed case ($\epsilon'=0$).
}
\label{fig_Xf_Ef_DJ}
\end{figure*}

\textbf{\textit{Case $\boldsymbol{j=1}$.}}
The equation~\ref{eq_e_fixed_E} becomes
\begin{equation}
X^3 = \left( \frac{2 \Delta J}{m-1} \right) X + \epsilon'.
\label{eq_roots_j_1}
\end{equation}
This cubic equation can be solved as described in Appendix~\ref{app_cubic_equation}. In particular, Eq.~\ref{eq_roots_j_1} is identical to Eq.~\ref{eq_depressed_cubic}, taking
\begin{equation}
\left\{
\begin{array}{l}
\displaystyle
p = -\frac{2 \Delta J}{m-1}, \\ \\
\displaystyle
q = -\epsilon'.
\end{array}
\right.
\label{eq_p_q_fixed_points}
\end{equation}
The discriminant of the cubic equation~\ref{eq_roots_j_1} is 
$$
\Delta = 32 \left( \frac{\Delta J}{m-1} \right)^3 - 27 \epsilon'^2
$$
For $\Delta < 0$, there is one fixed point given by Eq.~\ref{eq_one_real_solution}, and for 
$\Delta \geq 0$, there are three fixed points given by \ref{eq_three_real_solutions}.
These solutions are plotted in panel~(a) of Fig.~\ref{fig_Xf_Ef_DJ}.

\textbf{\textit{Case $\boldsymbol{j=2}$.}}
Eq.~\ref{eq_e_fixed_E} reads
\begin{equation}
E^2 = \left( \frac{4 \Delta J}{m-2} \right) + (-1)^k \epsilon'.
\label{eq_roots_j_2}
\end{equation}

For $k=0$ (resp. $k=1$), the fixed points are on the $X$-axis (resp. $Y$-axis).
The solutions of the equation above are
\begin{equation}
\begin{array}{l}
\displaystyle
X_{\rm f} =\pm \sqrt{ \left(\frac{4}{m-2} \right) \Delta J + \epsilon'} \\ \\
\displaystyle
Y_{\rm f} =\pm \sqrt{ \left(\frac{4}{m-2} \right) \Delta J - \epsilon'} \\
\end{array}
\end{equation}
and are plotted in panel~(b) of Fig.~\ref{fig_Xf_Ef_DJ}.

\textbf{\textit{Case $\boldsymbol{j=3}$.}}
Eq.~\ref{eq_e_fixed_E} provides
\begin{equation}
E^2 = \left( \frac{6 \Delta J}{m-3} \right) + (-1)^k \epsilon' E.
\label{eq_roots_j_3}
\end{equation}
The resulting solutions
\begin{equation}
\begin{array}{l}
\displaystyle
E_{\rm f} =  (-1)^k \left( \frac{\epsilon'}{2} \right) \pm \sqrt{ \left( \frac{6\Delta J}{m-3} \right) + \frac{\epsilon'^2}{4}}
\end{array}
\end{equation}
are plotted in panel~(c) Fig.~\ref{fig_Xf_Ef_DJ}.
The case $k=0$ corresponds to the fixed points along the $X$-axis, 
while $k=1$ corresponds to the fixed points along the $\pi/3$-axis.

\textbf{\textit{Case $\boldsymbol{j=4}$.}}
We now have
\begin{equation}
E^2 = \left( \frac{8 \Delta J}{m-4} \right) + (-1)^k \epsilon' E^2,
\label{eq_roots_j_4}
\end{equation}
which yields
\begin{equation}
\begin{array}{l}
\displaystyle
E_{\rm f} = \pm \sqrt{ \left( \frac{8 \Delta J}{m-4} \right) \left[ \frac{1}{ 1 - (-1)^k \epsilon'} \right] },
\end{array}
\end{equation}
see panel~(d) of Fig.~\ref{fig_Xf_Ef_DJ}.
The case $k=0$ corresponds to the fixed points along the $X$-axis, 
while $k=1$ corresponds to the fixed points along the $\pi/4$-axis.

\textbf{\textit{Case $\boldsymbol{j \geq 5}$.}}
A new regime appears beyond the order four.
The term containing $\epsilon'$ in Eq.~\ref{eq_e_fixed_E} 
is of order larger than two in $E$. 
Consequently, considering that both $\epsilon'$ and $E$ are small, we have
\begin{equation}
\displaystyle
E_{\rm f} \approx
\pm \sqrt{ \left( \frac{2j}{m-j} \right) \Delta J},
\label{eq_roots_j_geq_5}
\end{equation}
with a relative error of order $\epsilon' E_{\rm f}^{j-4}$.
These solutions are plotted in panel~(e) of Fig.~\ref{fig_Xf_Ef_DJ}.
The values of $E_{\rm f}$ are now independent of $\epsilon'$.
This means than the fixed points (excluding the origin) are distributed along a circle,
with $j$ elliptic points alternating with $j$ hyperbolic points.

Eq.~\ref{eq_roots_j_geq_5} has an straightforward interpretation. 
The expression of $\Delta J$ (Eq.~\ref{eq_definition_Delta_J}) implies that 
the fixed point corresponds to $a = a_0$.
This merely means that the corresponding orbits are then at exact resonance, as expected.
This is why the plot in panel~(e) of Fig.~\ref{fig_Xf_Ef_DJ} is undistinguishable from the unperturbed case
($\epsilon'=0$).

\section{Behavior of the eccentricity near a resonance}
\label{sec_J_e}

The behavior of ring particles in a dense collisional disk at the vicinity of a resonance is complex due to the combination of various effects, among which differential precession rate, self-gravity and viscous effects that lead to local angular momentum flux reversal.
These issues are best tackled using the equations of hydrodynamic or $N$-body collisional simulations, see Paper~II.

Meanwhile, it is instructive to estimate the limit superior $e_{\rm max}$ of the orbital eccentricity 
of test particles initially on circular orbits, knowing that collisions with tend to damp eccentricities below this value.
The Sections~\ref{sec_stability_1st_kind} and \ref{sec_2nd_kind_orbits} and 
Figs.~\ref{fig_maps_hamiltonians} and \ref{fig_Xf_Ef_DJ}
show that only three types of resonances yield unstable first-kind orbits:
$(i)$ 1$^{\rm st}$-order resonances, which force an eccentricity for any values of $\Delta J$;
$(ii)$ 2$^{\rm nd}$-order resonances, which force an eccentricity only inside a finite interval 
of $\Delta J$ (Eq.~\ref{eq_interval_DJ}); and
$(iii)$ third-order resonances, which force a non-zero eccentricity at the isolated value $\Delta J=0$.

\subsection{First-order resonances}

We consider a particle starting on a circular orbit with a Hamiltonian value ${\cal H}(0,0)={\cal H}_0$.
The particle then follows the level curve ${\cal H}(X,Y)= {\cal H}_0$.
The eccentricity reaches its maximum value $e_{\rm max} = |X_{\rm sup}|$ on the $X$-axis,
where $X_{\rm sup}$ is the non-zero solution of ${\cal H}(X,0)= {\cal H}_0$, i.e.
\begin{equation}
X^3 - \left( \frac{4\Delta J}{m-1} \right) X - 4\epsilon' = 0.
\label{eq_e_sup_first_order}
\end{equation}
Similarly to what was done in Section~\ref{sec_2nd_kind_orbits}, we identify Eq.~\ref{eq_e_sup_first_order} with the cubic equation~\ref{eq_depressed_cubic}, taking
\begin{equation}
\left\{
\begin{array}{l}
\displaystyle
p = -\frac{4 \Delta J}{m-1}, \\ \\
\displaystyle
q = -4\epsilon',
\end{array}
\right.
\label{eq_p_q_discontinuity}
\end{equation}
from which we obtain the discriminant
$$
\Delta = 16 \left[ 16 \left( \frac{\Delta J}{m-1} \right)^3 - 27 \epsilon'^2 \right].
$$

In the case of $\Delta < 0$, the solution is, from Eq.~\ref{eq_one_real_solution},
\begin{equation}
e_{\rm max}= 
\left| \left( \frac{-q - \sqrt{-\Delta/27}}{2} \right)^{1/3} + \left( \frac{-q + \sqrt{-\Delta/27}}{2} \right)^{1/3} \right|.
\label{eq_emax_j_1_branch_1}
\end{equation}

If $\Delta \geq 0$, there are three possible solutions given by Eq.~\ref{eq_three_real_solutions}.
The one we are looking for is the closest to the origin, due to collision damping.
Considerations on the arguments of the cosine functions
in Eq.~\ref{eq_three_real_solutions} show that it corresponds to the case $k=2$, i.e.
\begin{equation}
e_{\rm max}= 
\left| 2\sqrt{\frac{-p}{3}} \cos \left[ \frac{1}{3} \arccos \left( \frac{3q}{2p} \sqrt{\frac{-3}{p}} \right) + \frac{4\pi}{3} \right] \right|.
\label{eq_emax_j_1_branch_2}
\end{equation}

The equations~\ref{eq_emax_j_1_branch_1} and \ref{eq_emax_j_1_branch_2} 
define two branches with a discontinuity at $\Delta= 0$, i.e. at
$\Delta \overline{a}/a_0 = (3/2)(m-1)|2\epsilon'|^{2/3}$.
At that value, $e_{\rm max}$ suffers a discontinuity and 
jumps from $|2\epsilon'|^{1/3}$ to $|16\epsilon'|^{1/3}$.
Finally, from Eq.~\ref{eq_e_sup_first_order}, we note that for 
$\Delta J =0$, i.e. $\overline{a}=0$, we have $e_{\rm max}= |4\epsilon'|^{1/3}$.

\begin{figure}[t!]
\centerline{\includegraphics[totalheight=42mm,trim=0 0 0 0]{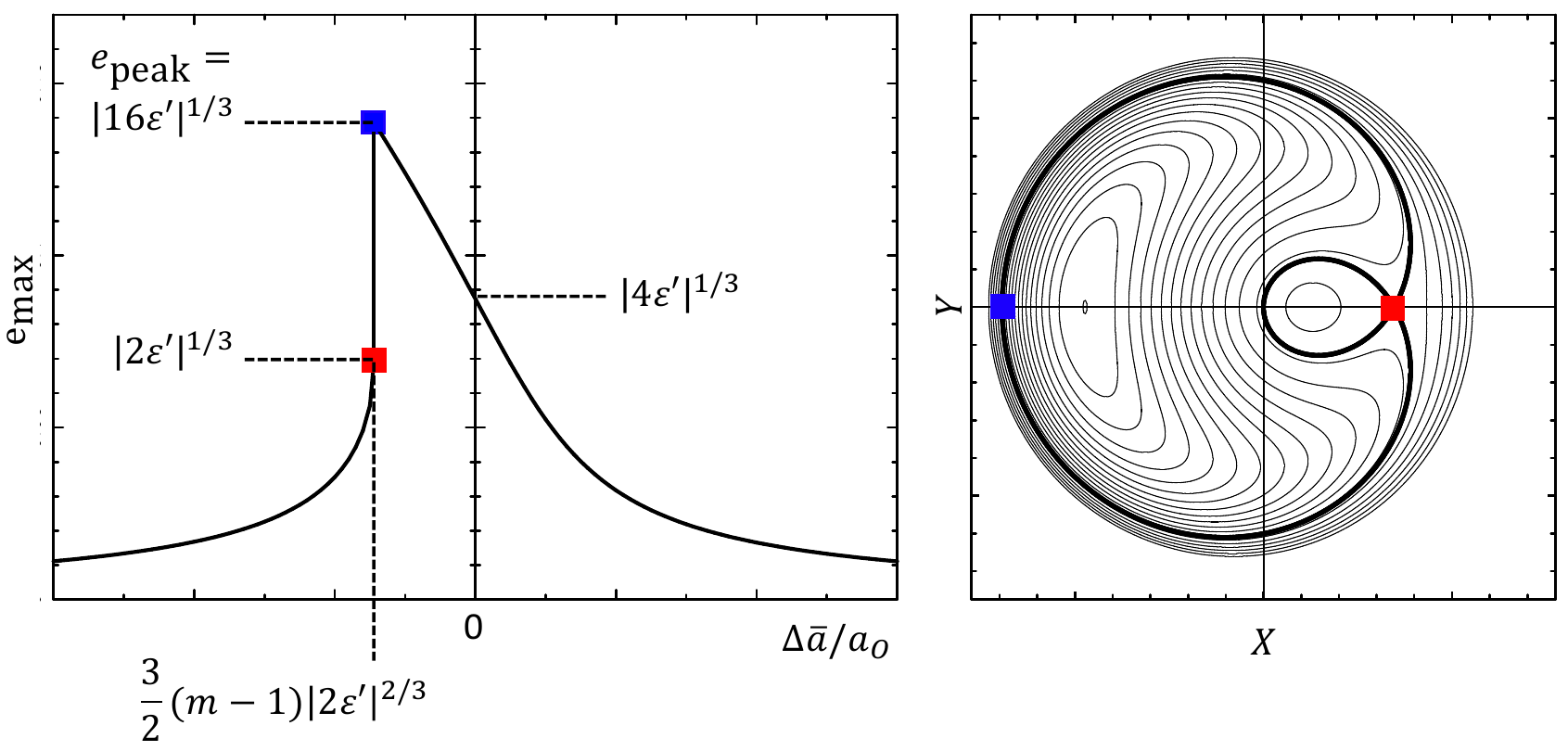}}
\caption{
Response to a 1$^{\rm st}$-order resonance. 
\textit{Left panel}:
the maximum eccentricity $e_{\rm max}$ reached by a particle initially on a circular orbit 
with modified semi-major axis $\Delta \overline{a}/a_0$, 
with $m < 0$ and $\epsilon' < 0$.
The right (resp. left) branch of the function is given by Eq.~\ref{eq_emax_j_1_branch_1} (resp. \ref{eq_emax_j_1_branch_2}).
The value of $e_{\rm max}$ suffers a discontinuity at $\Delta \overline{a}/a_0 = (3/2)(m-1)|2\epsilon'|^{2/3}$, 
where $e_{\rm max}$ jumps from  $|2\epsilon'|^{1/3}$ to $|16\epsilon'|^{1/3}$,
the maximum possible eccentricity $e_{\rm peak}$.
\textit{Right panel}: the phase portrait corresponding to the discontinuity, 
with the homoclinic trajectory going through the origin.
The red and blue points correspond to their counterparts shown in the left panel.
}
\label{fig_DJ_emax_1st_order}
\end{figure}

The general variation of $e_{\rm max}$ with $\overline{a}$
is displayed in Fig.~\ref{fig_DJ_emax_1st_order}, 
where $e_{\rm max}$ is plotted as a function of the distance 
$\Delta \overline{a}/a_0$ to exact resonance.
In this figure, a particle moves on the average vertically since $\overline{a}$ is conserved.
If the particle starts on the horizontal axis, i.e. with $e=0$, it moves up vertically (because $\overline{a}$ is conserved) to the bell-shaped curve defined by $e_{\rm max}$, and then returns to the horizontal axis.
From this figure, we can define the width $W$ in $\overline{a}$ and the peak value $e_{\rm peak}$ of $e_{\rm max}(\overline{a})$ as, 
\begin{equation}
\begin{array}{ll}
W \sim 3|m-1||2\epsilon'|^{2/3} a_0 & {\rm and~~~} e_{\rm peak} = |16\epsilon'|^{1/3}.
\end{array}
\label{eq_emax_max_width_1}
\end{equation}
More precisely, $W$ is defined as twice the distance of the discontinuity of $e_{\rm max}$ to the 
origin $\overline{a}=0$. While this is somehow arbitrary, this offers an estimate of the span in
$\overline{a}$ where $e_{\rm max}$ is significant.
Because $e$ is usually small, $W$ is also a good estimation of the span in semi-major axis over which initially circular orbits acquire a significant eccentricity.
This definition of the width $W$ is non-standard when compared to definitions given in classical text books (e.g. \citealt{murray2000}), that is the maximum variation in semi-major axis of an orbit with librating resonant angle $\phi$.

Our definition of $W$, however, is more useful in the context of dense collisional rings. When plotted in Fig.~\ref{fig_DJ_emax_1st_order} and Fig.~\ref{fig_DJ_emax_2nd_order}), a ring particle tends to move vertically in the region of the bell-shaped curve, due to the resonance forcing. Conversely, collisions will tend to push the particle down the horizontal axis due to eccentricity damping, until a stationary regime is reached. This behavior is analyzed in the simulations presented in Paper~II. 

An important parameter is the time scale necessary to build up the eccentricity from zero to its maximum value $e_{\rm max}$. As an example, we consider a particle initially on a circular orbit at exact resonance ($\Delta \overline{a}=0$). The equations of motions~\ref{eq_motion_X_Y} provide the rate of change of the eccentricity near the origin $X$=$Y$=0. Considering that the maximum eccentricity reached by this particle is $|4\epsilon'|^{1/3}$ (Fig.~\ref{fig_DJ_emax_1st_order}), it can be shown that the excitation time scale at exact resonance is
\begin{equation}
T_{\rm res} \approx \frac{2}{3\pi |m(m-1)|} \left( \frac{1}{2\epsilon'^2} \right)^{1/3} T_{\rm cor},
\label{eq_Tres_1st_order}
\end{equation}
where $T_{\rm cor}$ is the orbital period at corotation, and thus also the rotation period of the body.

\subsection{Second-order resonances}

\begin{figure}[t!]
\centerline{\includegraphics[totalheight=42mm,trim=0 0 0 0]{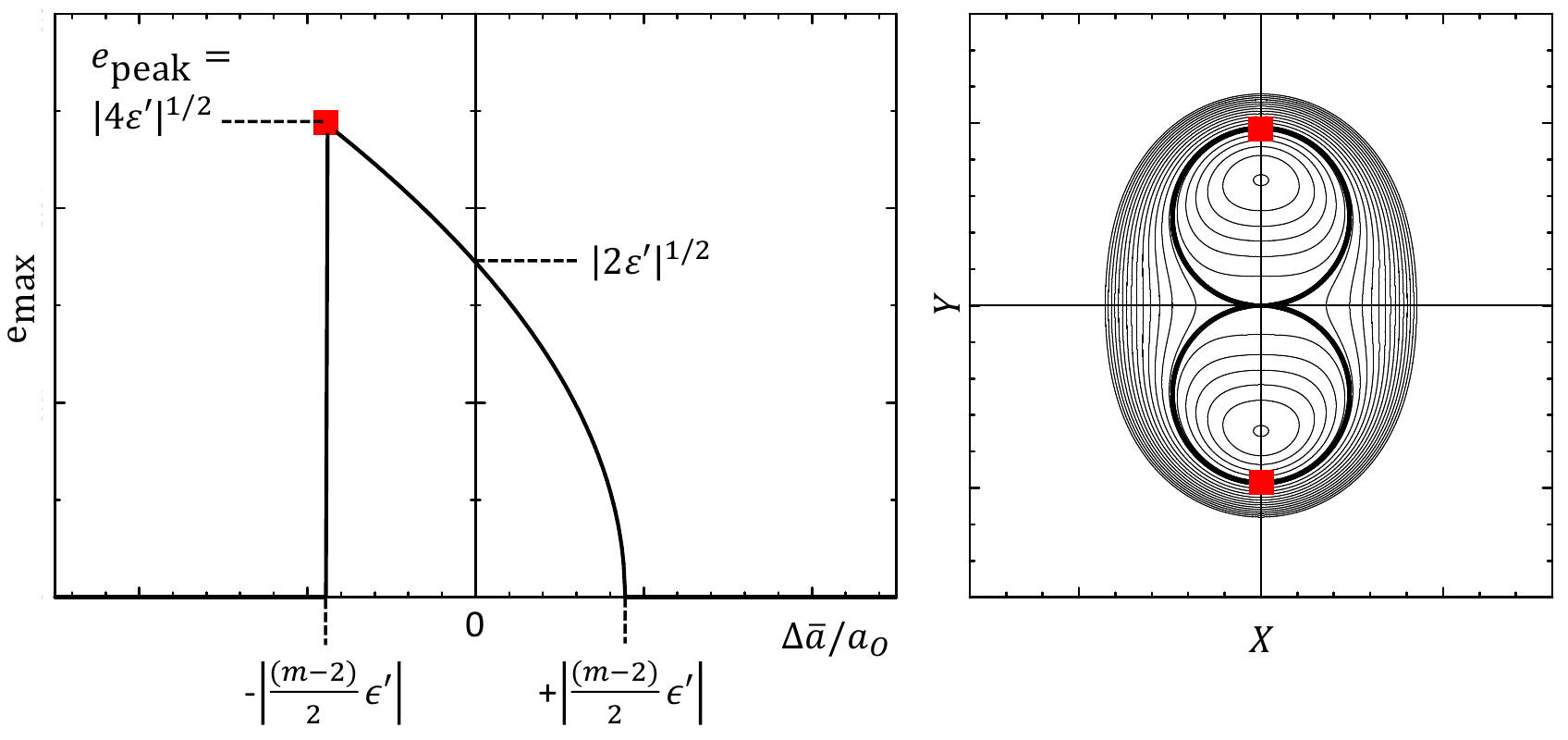}}
\caption{
The same as Fig.~\ref{fig_DJ_emax_1st_order} for a 2$^{\rm nd}$-order resonance. 
\textit{Left panel}:
the function plotted here is given by Eq.~\ref{eq_emax_j_2}.
The value of $e_{\rm max}$ suffers a discontinuity at $\Delta \overline{a}/a_0 = -|(m-2)\epsilon'/2|$, 
where  $e_{\rm max}$ jumps from zero to is maximum value $e_{\rm peak}= |4\epsilon'|^{1/2}$.
\textit{Right panel}: the phase portrait corresponding to the discontinuity. 
The red points correspond to their counterpart of the left panel.
}
\label{fig_DJ_emax_2nd_order}
\end{figure}

For $|\Delta \overline{a}/a_0| > |(m-2)\epsilon'/2|$, the origin of the phase portrait 
is a stable elliptic point (Section~\ref{sec_stability_1st_kind} and Fig.~\ref{fig_Xf_Ef_DJ}), 
so that $e_{\rm max}=0$ in this domain.
Conversely, for $|\Delta \overline{a}/a_0| \leq |(m-2)\epsilon'/2|$, 
the level curve going through the origin is a 8-shaped curve
defined by ${\cal H}(X,Y)= 0$ (Fig.~\ref{fig_maps_hamiltonians}). 
Taking $j=2$ in Eq.~\ref{eq_hamil_expanded_in_e}, we obtain
\begin{equation}
\begin{array}{ll}
\displaystyle e_{\rm max}= 
\sqrt{2|\epsilon'| + \left( \frac{4}{m-2} \right) \frac{\Delta \overline{a}}{a_0}} &
\displaystyle {\rm for~~}
\left| \frac{\Delta \overline{a}}{a_0} \right| \leq \left| \frac{(m-2)\epsilon'}{2} \right|, \\ \\
\displaystyle e_{\rm max}= 0 & 
\displaystyle {\rm for~~}
\left| \frac{\Delta \overline{a}}{a_0} \right| > \left| \frac{(m-2)\epsilon'}{2} \right|. 
\end{array}
\label{eq_emax_j_2}
\end{equation}
The value of $e_{\rm max}$ as a function of $\Delta \overline{a}/a_0$ is plotted 
in Fig.~\ref{fig_DJ_emax_2nd_order} (left panel), 
with a discontinuity at  $\Delta \overline{a}/a_0 = -|(m-2)\epsilon'/2|$.
The phase portrait for that value is displayed in the right panel of Fig.~\ref{fig_DJ_emax_2nd_order}.

The width over which $e_{\rm max}$ is non-zero and the value $e_{\rm peak}$ are now 
\begin{equation}
\begin{array}{ll}
W = |(m-2)\epsilon'| a_0 & {\rm and~~~} e_{\rm peak} = |4\epsilon'|^{1/2}.
\end{array}
\label{eq_emax_max_width_2}
\end{equation}

The same exercise as for 1$^{\rm st}$-order resonance provides the time scale $T_{\rm res}$ for building up the orbital eccentricity of a particle starting on a circular orbit with $\overline{a}=0$. The origin of the phase portrait being a saddle point (Fig.~\ref{fig_maps_hamiltonians}), the particle moves away from this origin exponentially. Using again the equations of motions~\ref{eq_motion_X_Y}, it can be shown that the $e$-folding time scale for the growth of eccentricity is
\begin{equation}
T_{\rm res} \approx \frac{4}{3\pi |m(m-2)\epsilon'|} T_{\rm cor}.
\label{eq_Tres_2nd_order}
\end{equation}

\subsection{Third-order resonances}

\begin{figure}[t!]
\centerline{\includegraphics[totalheight=38mm,trim=0 0 0 0]{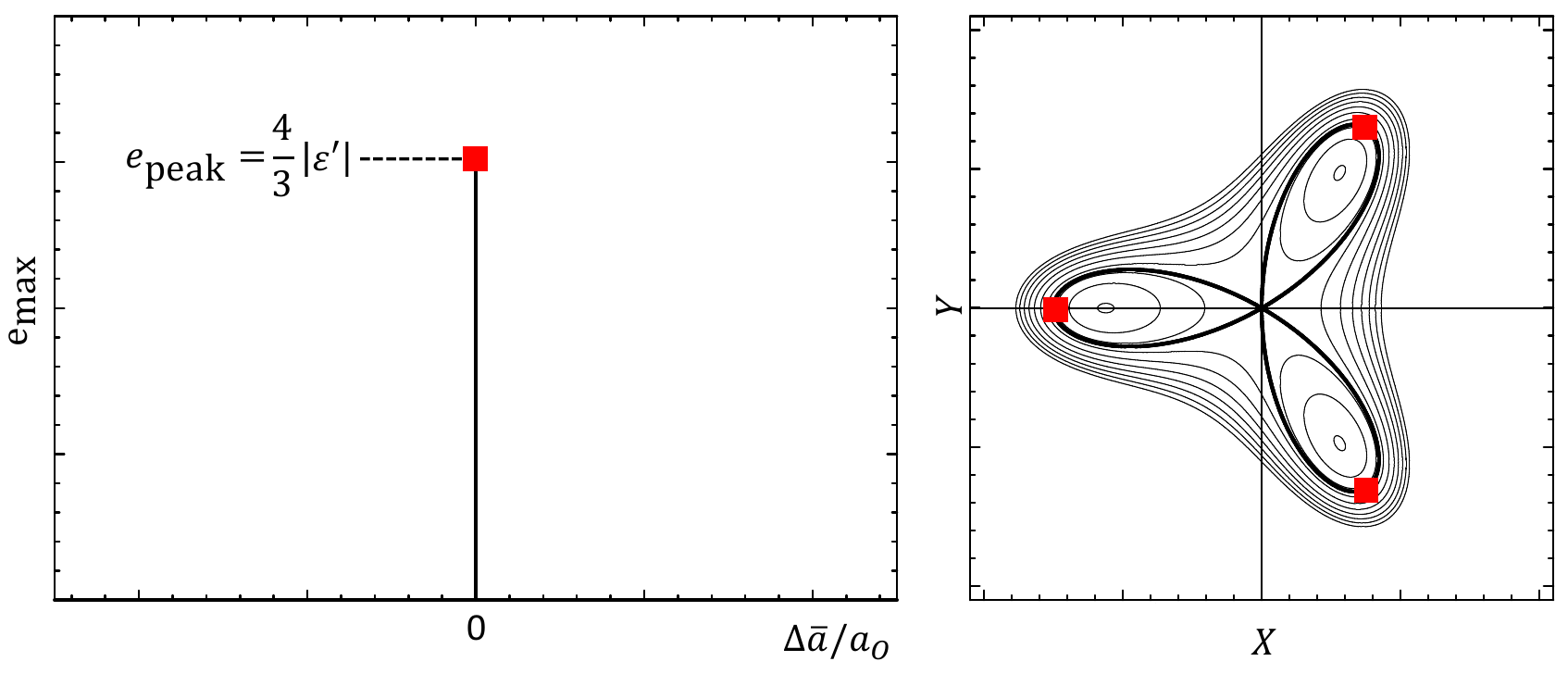}}
\caption{
The same as Fig.~\ref{fig_DJ_emax_2nd_order} for a third-order resonance. 
\textit{Left panel}:
the function plotted here is given by Eq.~\ref{eq_emax_j_3}.
\textit{Right panel}: the phase portrait corresponding to the discontinuity at $\Delta \overline{a}/a_0= 0$. 
The red points correspond to their counterpart of the left panel.
}
\label{fig_DJ_emax_3rd_order}
\end{figure}

The origin is an unstable hyperbolic point only for the isolated value $\Delta J= \Delta \overline{a}/a_0= 0$.
The equation~\ref{eq_hamil_expanded_in_e} then provides $e_{\rm max}$ through the equation
$$
-\frac{3}{4} e^4 + \epsilon' e^3 \cos(3\phi)= 0.
$$ 
This yields,
\begin{equation}
\left\{
\begin{array}{ll}
\displaystyle e_{\rm max}= \frac{4}{3} |\epsilon'| &
\displaystyle {\rm for~~}
\frac{\Delta \overline{a}}{a_0} = 0, \\ \\
\displaystyle e_{\rm max}= 0 & 
\displaystyle {\rm for~~}
\frac{\Delta \overline{a}}{a_0} \neq 0,
\end{array}
\right.
\label{eq_emax_j_3}
\end{equation}
see Fig.~\ref{fig_DJ_emax_3rd_order}.
Thus, for third-order resonances, we have
\begin{equation}
\begin{array}{ll}
\displaystyle
W = 0 & {\rm and~~~} e_{\rm peak} = \frac{4}{3} |\epsilon'|.
\end{array}
\label{eq_emax_max_width_3}
\end{equation}

We do not estimate here the resonant excitation time $T_{\rm res}$ for third-order resonances, as their width is zero, so that colliding particles cannot stay at exact resonance during the excitation process.

The values of $W$, $e_{\rm peak}$ and resonant excitation time $T_{\rm res}$ obtained for first-, second- and third-order resonances are summarized in Table~\ref{tab_resonance_W_emax}.
This table also provides the dependence of these quantities with respect to the mass anomaly $\mu$ and the elongation parameter $C_{22}$ of the body.

\begin{table*}[!t]
\centering
\caption{
Resonance widths, maximum eccentricities and resonant excitation time.
\label{tab_resonance_W_emax}
}
\renewcommand{\arraystretch}{1.5}
\begin{tabular}{llll}
\hline \hline
Resonance 	    & $W/a_0$ 						& $e_{\rm peak}$           & $T_{\rm res}/T_{\rm cor}$ \\
\hline
First-order    	&  $3|(m-1)||2\epsilon'|^{2/3}$	& $|16\epsilon'|^{1/3}$   & $2/|3\pi m(m-1) (2\epsilon'^2)^{1/3}|$ \\
Second-order 	& $|(m-2)\epsilon'|$ 			& $|4\epsilon'|^{1/2}$    & $4/|3\pi m(m-2) \epsilon'|$ \\
Third-order 	& 0 							& $(4/3) |\epsilon'|$     & NA \\
\hline
\multicolumn{4}{c}{Dependence on $\mu$ and $C_{22}$} \\
\hline
First-order  & $\propto \mu^{2/3}$, $\propto C_{22}^{|m|/3}$ & $\propto \mu^{1/3}$, $\propto  C_{22}^{|m|/6}$ & $\propto  \mu^{-2/3}$, $\propto C_{22}^{-|m|/3}$ \\
Second-order & $\propto \mu$, $\propto C_{22}^{|m|/2}$        & $\propto \mu^{1/2}$, $\propto C_{22}^{|m|/4}$ & $\propto \mu^{-1}$, $\propto  C_{22}^{-|m|/2}$  \\
Third-order  & NA                             & $\propto \mu$, $\propto C_{22}^{|m|/2}$       & NA \\
\hline
\end{tabular}
\end{table*}

\section{Resonance order and orbit structure}
\label{sec_order_and_self_crossing}

The response of a collisional disk to a SOR depends on two criteria:
$(i)$ the order of the resonance, which sets the typical eccentricities and the interval
of $\overline{a}$ over which a significant response of the disk is expected
(Figs.~\ref{fig_DJ_emax_1st_order} and \ref{fig_DJ_emax_2nd_order});
$(ii)$ The structure of the periodic resonant orbits near the resonance, in particular
the possible presence of self-intersecting points along these orbits,
as observed in a frame rotating with the body.

\begin{figure}
\centerline{
\includegraphics[totalheight=65mm,trim=0 0 0 0]{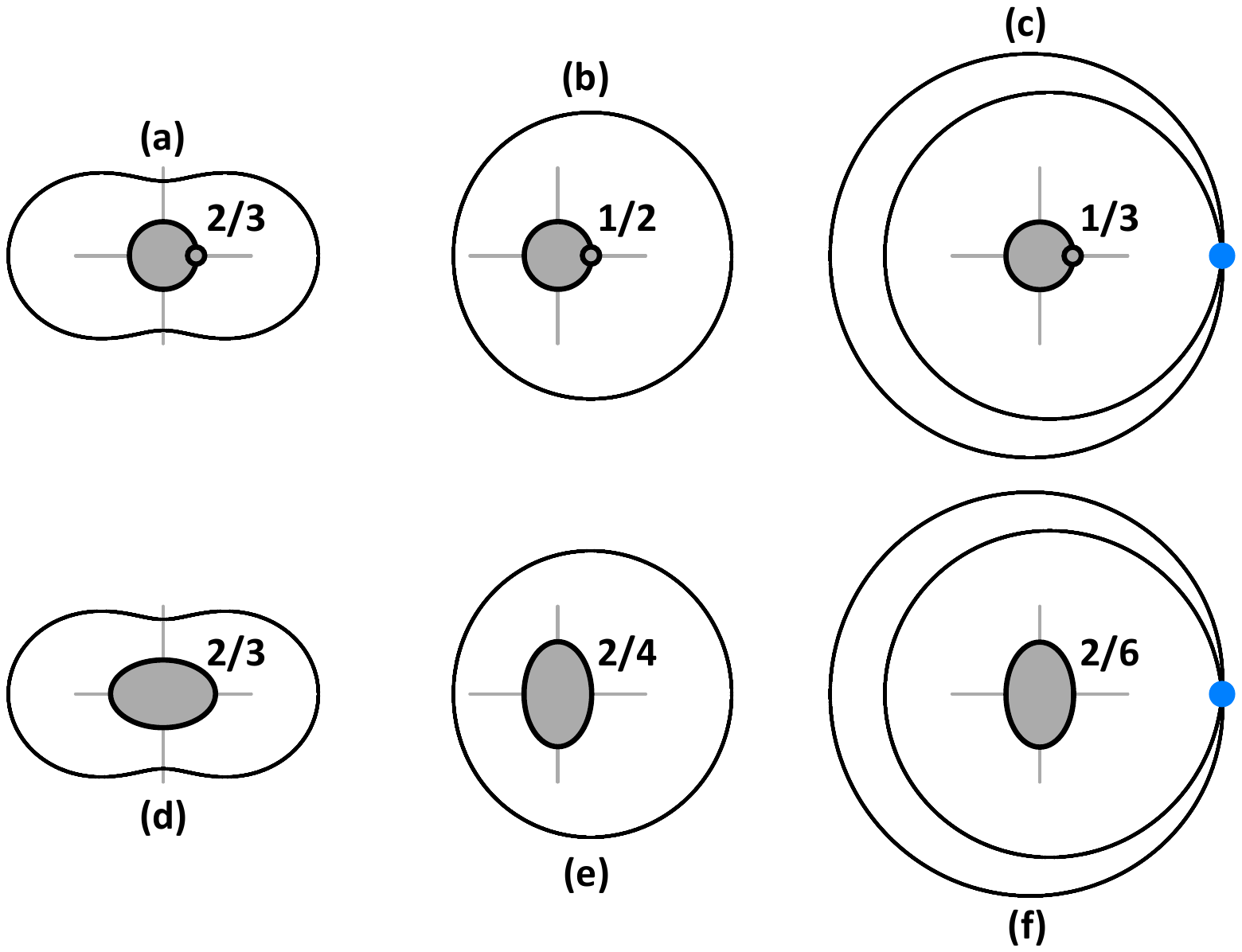}
}
\caption{
Examples of resonant periodic orbits. 
\textit{Panels~(a), (b) and (c)}: 
case of a body with a mass anomaly, observed in a frame rotating with the body.
\textit{Panels~(d), (e) and (f)}: the same around a triaxial body.
The orbits (a) and (d) have the same structure and correspond to the same order of resonance.
The orbits (b) and (e) have the same structure but correspond to different resonance orders
(one and two, respectively).
The orbits (c) and (f) have the same structure with one self-intersection point (blue dot)
and are associated with resonances of orders two and four, respectively.
}
\label{fig_periodic_orbits}
\end{figure}

This structure is entirely defined by the ratio $n/\Omega_{\rm B} \approx m/(m-j)$ (Eq.~\ref{eq_ratio_n_omega_exact}).
In particular, a resonant periodic orbit has $|m'|(j'-1)$ self-intersecting points, where $m'$ and $j'$ are 
the relatively prime versions of $m$ and $j$ \citep{sicardy2020a,sicardy2020b}.
In a collisional disk, this implies that the resonant streamlines forced near a $m/(m-j)$ SOR have $|m'|(j'-1)$ self-crossings points. Thus, only the Lindblad resonances ($j=j'=1$) avoid the self-crossing problem (Fig.~\ref{fig_periodic_orbits}). This allows analytical solutions to be derived, with nested periodic neighboring orbits that interact to create spiral features. From this formalism, the description of angular momentum transfer and confinement mechanisms is possible.
For $j' \geq 2$, a resonant streamline has  at least one self-crossing point, where the density and velocity shear become undefined. A study of these cases requires numerical simulations, the topic of Paper~II. 

For higher-order resonances, the number of self-intersecting points
of a periodic orbit depends on whether $m$ and $j$ are relatively primes.
For instance in the case of a triaxial body, only even values of $m$ are allowed from the symmetry
of the potential (Eq.~\ref{eq_pot_ell}).
Thus, the resonance $n/\Omega_{\rm B} \approx 1/2$ is in fact a $n/\Omega_{\rm B} \approx 2/4$ 2$^{\rm nd}$-order resonance
with $m$=-2 and $j=2$.
Even though the periodic orbit looks like that of a 1$^{\rm st}$-order resonance
(in particular it has no self-intersection since $m'$=-1 and $j'$=1), 
it is actually a 2$^{\rm nd}$-order resonant orbit, and as such will behave as shown in 
Fig.~\ref{fig_DJ_emax_2nd_order}. 
 
Similarly, the $n/\Omega_{\rm B} \approx 1/3$ resonance around a triaxial body is in fact a $n/\Omega_{\rm B} \approx 2/6$ fourth-order resonance, but now the periodic orbits have one self-intersecting point since $m'=-1$ and $j'=2$ (Fig.~\ref{fig_periodic_orbits}).

In summary, the structure of a resonant orbit alone is not sufficient to infer the order of the resonance. The order also depends on the symmetry of the potential at the origin of this resonance. 

\section{Applications to resonances around Chariklo, Haumea and Quaoar}
\label{sec_resonances_cha_haum_qua}

We now apply our results to Chariklo, Haumea and Quaoar.
Only 1$^{\rm st}$- and 2$^{\rm nd}$-order resonances are considered, as they are the only ones that excite the
orbital eccentricity $e$ of an initially circular orbit over a finite interval of $\overline{a}$, see Figs.~\ref{fig_DJ_emax_1st_order} and \ref{fig_DJ_emax_2nd_order}.

Two types of non-axisymmetric potentials are considered in this paper: 
a triaxial body which creates a quadrupole potential and 
a mass anomaly which creates a dipole-type potential.

\begin{figure}[!h]
\centerline{\includegraphics[totalheight=58mm,trim=0 0 0 0]{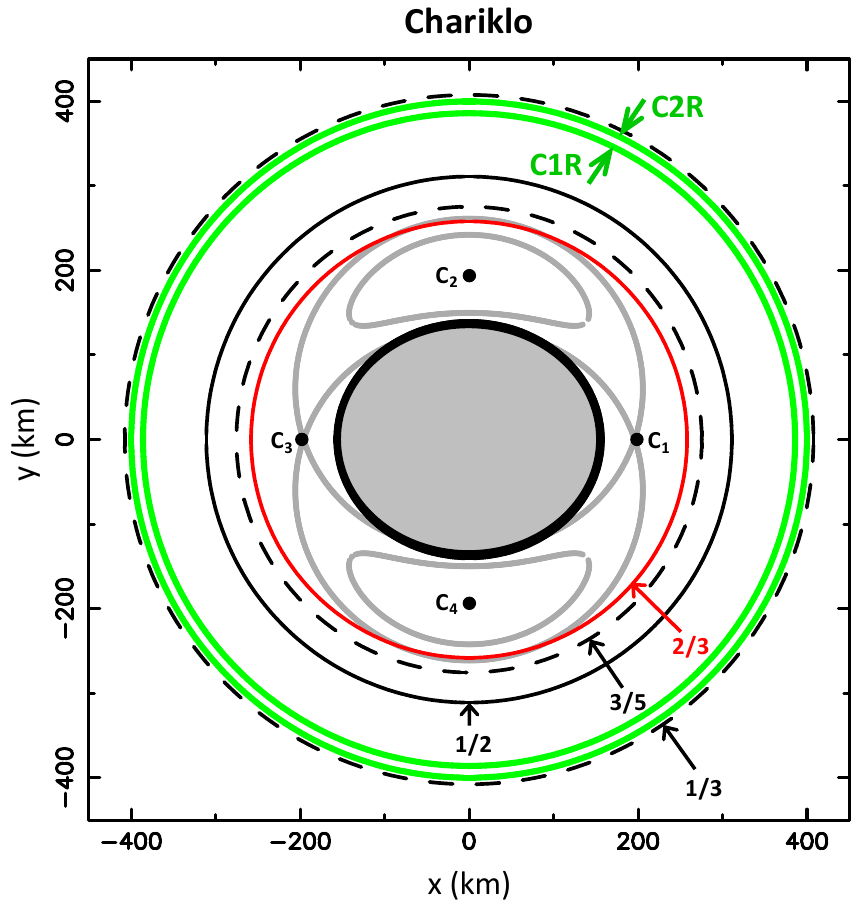}}
\centerline{\includegraphics[totalheight=55mm,trim=0 0 0 0]{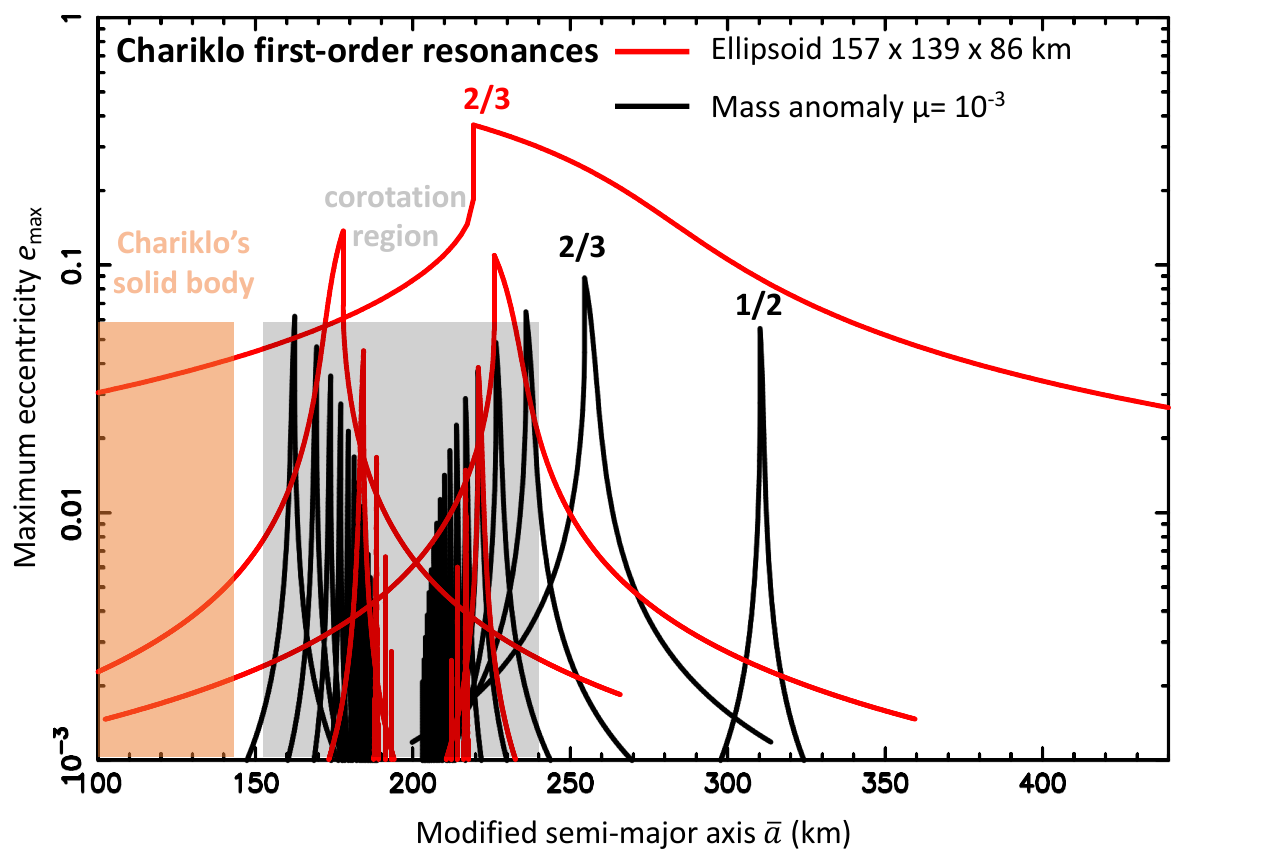}}
\centerline{\includegraphics[totalheight=55mm,trim=0 0 0 0]{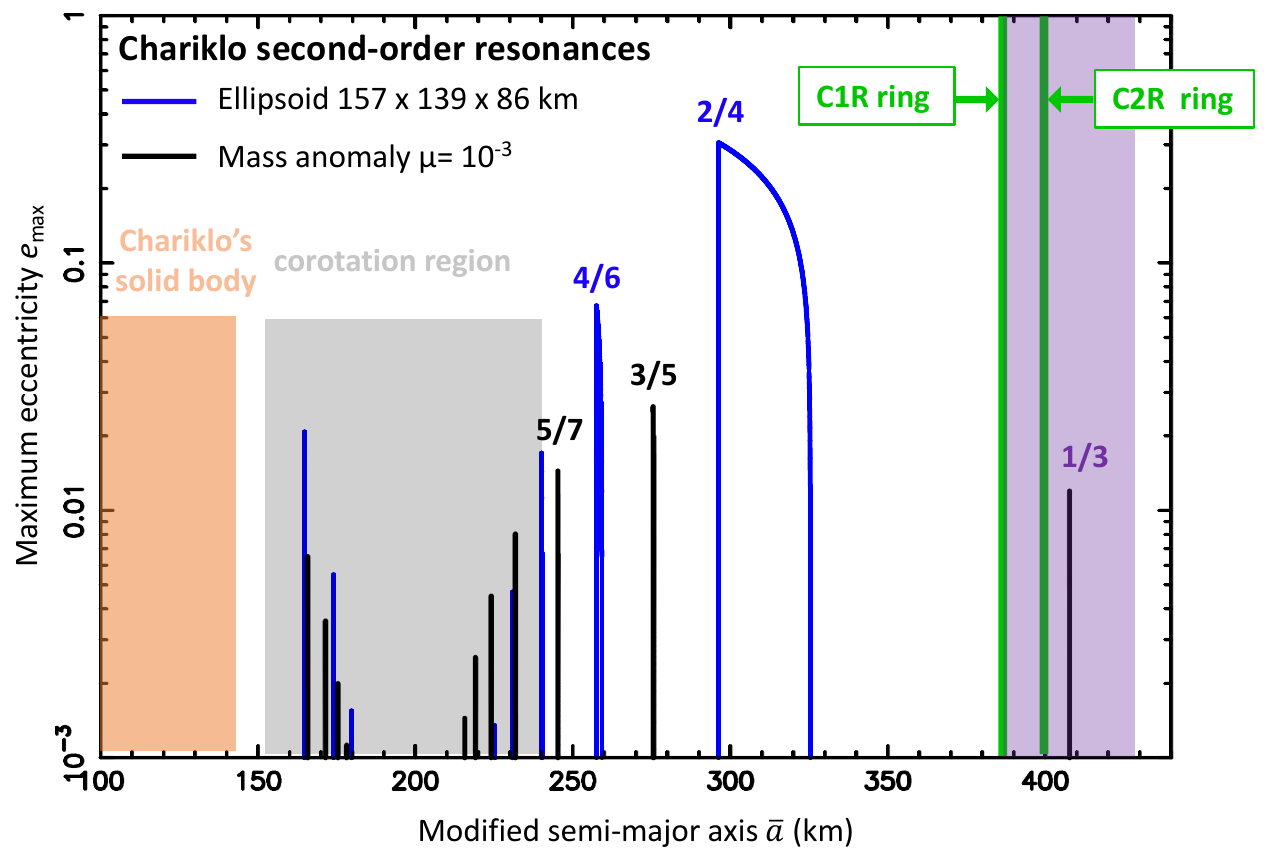}}
\caption{
Resonances around Chariklo.
\textit{Upper panel}:
the light gray lines show trajectories around the corotation points $C_2$ and $C_4$ of Chariklo, using Eq.~\ref{eq_trajec_corot}. 
The dark gray ellipse is a pole-on view of Chariklo's shape, taken from Table~\ref{tab_param_cha_hau_qua}. 
The two green circles mark the radii of C1R and C2R rings.
Red circle: the first-order 2/3 resonance caused by Chariklo's triaxial shape;
Solid black circle: the first-order 1/2 resonances caused by a mass anomaly;
Dashed black circle: the second-order 1/3 and 3/5 resonances caused by a mass anomaly.
More resonances radii are plotted in the lower panels.
\textit{Middle panel}: 
The maximum eccentricity $e_{\rm max}$ (in log-scale) reached by a particle initially on a circular orbit,
reproducing the behavior displayed in Fig.~ \ref{fig_DJ_emax_1st_order} for each SOR.
Red curves: the SORs caused by the triaxial shape of Chariklo.
Black curves: the SORs caused by a mass anomaly $\mu=10^{-3}$.
The orange box indicates Chariklo's largest semi-axis, while
the gray box shows the radial extension of the corotation zone, 
i.e. the full width of the corotation resonance (Eq.~\ref{eq_full_width_corot}).
\textit{Lower panel}: 
The same for second-order resonances. The radii of the rings C1R and C2R are marked in green.  The purple zone is the uncertainty on the 1/3 SOR location, due to the uncertainty on Chariklo's mass. The uncertainties on the ring radii are negligible at this scale, see Table~\ref{tab_param_cha_hau_qua}.
}
\label{fig_emax_vs_a_cha}
\end{figure}

\begin{figure}
\centerline{\includegraphics[totalheight=58mm,trim=0 0 0 0]{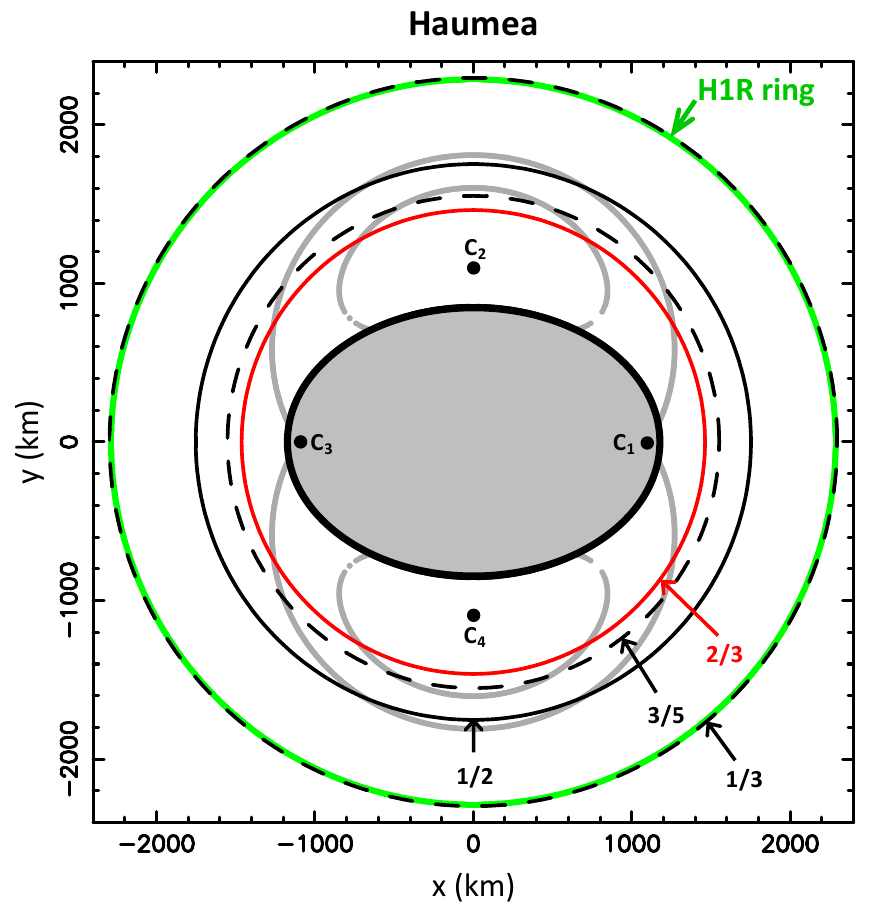}}
\centerline{\includegraphics[totalheight=55mm,trim=0 0 0 0]{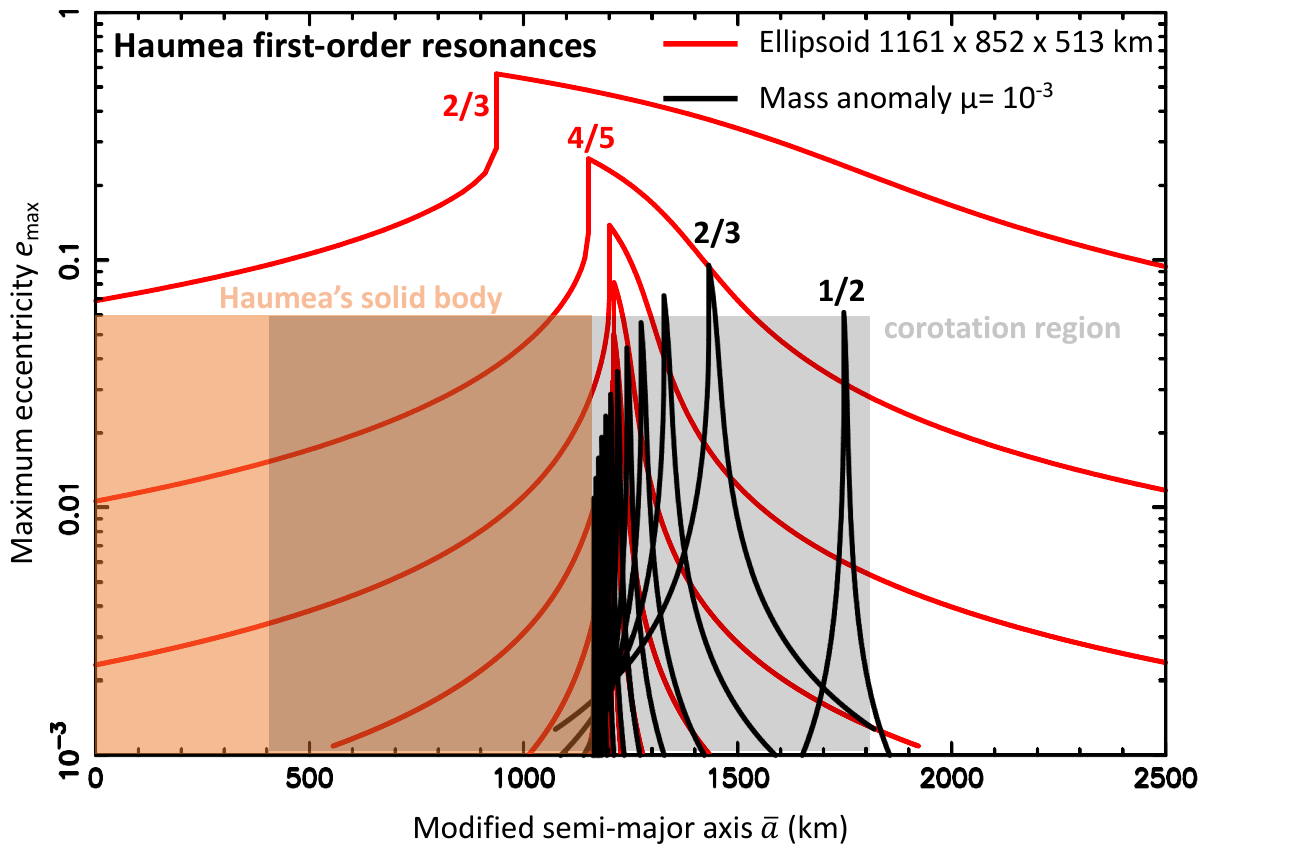}}
\centerline{\includegraphics[totalheight=55mm,trim=0 0 0 0]{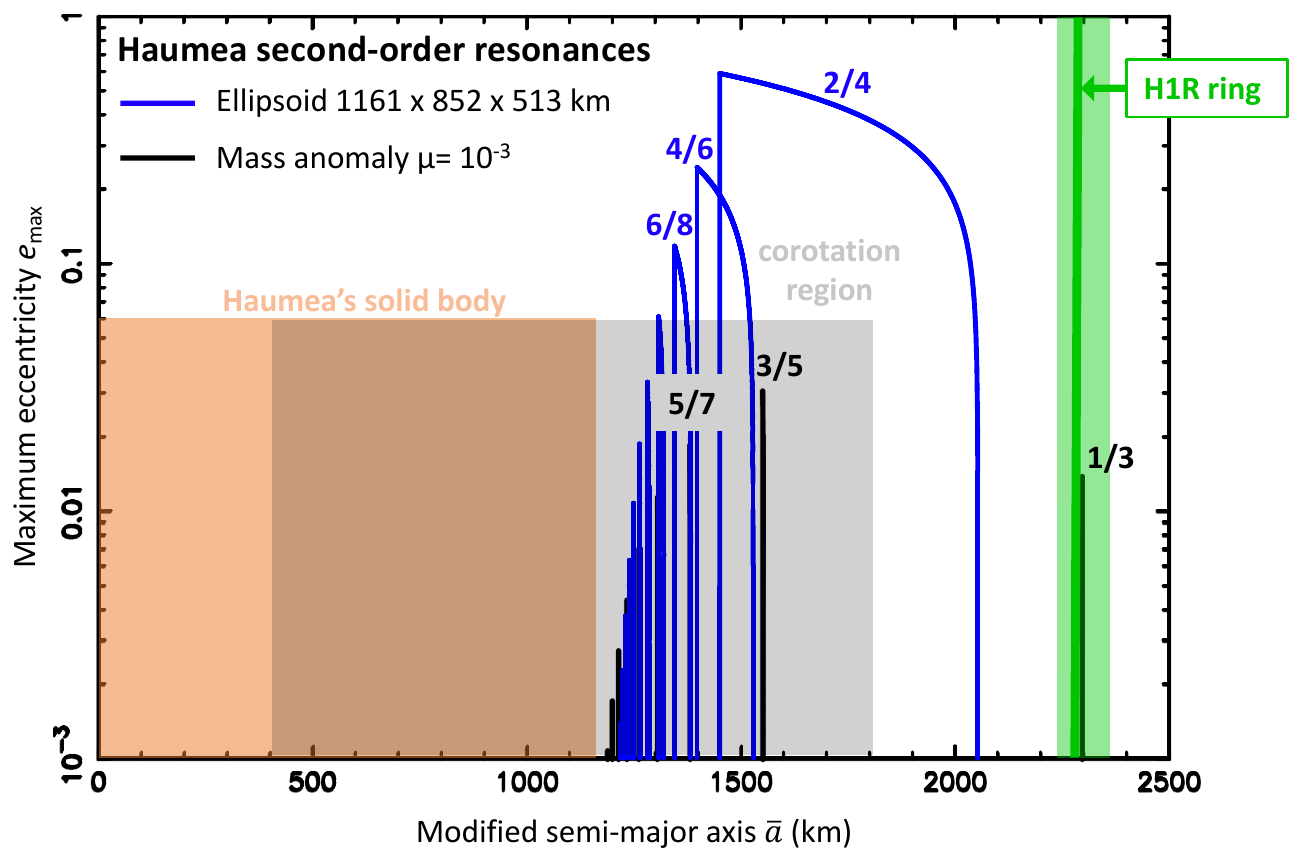}}
\caption{
The same as Fig.~\ref{fig_emax_vs_a_cha} for Haumea.
Here we use the nomenclature H1R to be in line with the names of
Chariklo's and Quaoar's rings (C1R, C2R, Q1R and Q2R).
The corotation zone now largely overlaps with the solid body.
The uncertainty on the ring radius in the right panel is indicated by the green box, 
while the uncertainty on the 1/3 resonance radius is negligible at this scale,
see Table~\ref{tab_param_cha_hau_qua}.
}
\label{fig_emax_vs_a_hau}
\end{figure}

The triaxial case assumes a homogeneous ellipsoid with principal semi-axes $A > B > C$,
from which the elongation $C_{22}$ is derived (see Appendix~\ref{app_potential_ellipsoid} for additional details).
The adopted physical parameters of Chariklo, Haumea and Quaoar for the ellipsoid case 
are listed in Table~\ref{tab_param_cha_hau_qua}
and have been used to generate Figs.~\ref{fig_emax_vs_a_cha}, \ref{fig_emax_vs_a_hau} and \ref{fig_emax_vs_a_qua},
showing a summary of the resonance and ring locations.

The mass anomaly case is described by a point-like ``mascon" of mass $\mu$ 
relative to the  body and located at the reference radius $R_{\rm ref}$ 
from the body center, see Table~\ref{tab_param_cha_hau_qua}.
No information is currently available for the values of $\mu$ concerning the three bodies. Here we adopt $\mu=10^{-3}$ as a guideline because it corresponds in order of magnitude to the value that permit the confinement of material near the 1/3 SOR, 
based on the simulations presented in Paper~II. As more information is gathered on Chariklo, Haumea and Quaoar, the estimation of $\mu$ can be refined and the values of $W$ and $e_{\rm peak}$ in Table~\ref{tab_resonance_W_emax} can be updated.

\begin{table*}[!t]
\centering
\caption{
Adopted physical parameters of Chariklo, Haumea and Quaoar\tablefootmark{1}.
\label{tab_param_cha_hau_qua}
}
\begin{tabular}{llll}
\hline \hline
& Chariklo\tablefootmark{2} & Haumea\tablefootmark{3}  & Quaoar\tablefootmark{4} \\
 \hline
Mass (kg) & $7 \times 10^{18}$           & $4.006 \times 10^{21}$ & $1.2 \times 10^{21}$ \\
Semi-axes $A \times B \times C$ (km)     & $157 \times 139 \times 86$ & $1161 \times 852 \times 513$ & $580 \times 513 \times 471$ \\
Reference radius $R_{\rm ref}$ (km)      & 115       & 712      &  516 \\
Elongation $C_{22}$                      & 0.0201    & 0.0614   & 0.0138 \\ 
Dynamical oblateness $J_2$               & 0.221     & 0.305    & 0.0586 \\ 
Rotation period (h)                      & 7.004     & 3.915341 & 17.6788 \\
Rotational parameter $q$                 & 0.202     & 0.268    &  0.0167 \\
Corotation radius (km)                   & 196       & 1104     & 2018 \\
Corotation full width $W_{\rm cor}$ (km) & 130        & 1410     & 484 \\

\hline
\multicolumn{4}{c}{Ring and resonance radii (km)} \\
\hline
Rings      & Q1R: $385.9\pm 0.4$  & H1R: $2287^{+75}_{-45}$ & Q1R: $4057 \pm 6$  \\
           & Q2R: $399.8\pm 0.6$  &                         & Q2R: $2520 \pm 20$ \\
Resonances & 1/3: $408 \pm 20$    & 1/3: $2285 \pm 8$       & 1/3: $4197 \pm 58$ \\
           &                      &                         & 5/7: $2525 \pm 35$ \\
\hline
\end{tabular}
\tablefoot{
\tablefoottext{1}{See definitions in Appendices~\ref{app_potential_mass_anomaly}, \ref{app_potential_ellipsoid},  and \ref{app_potential_corotation}.}
\tablefoottext{2}{\cite{leiva2017,morgado2021}.}
\tablefoottext{3}{\cite{ortiz2017}.}
\tablefoottext{4}{\cite{ortiz2003,vachier2012,morgado2023,pereira2023}}
}
\end{table*}

The figures~\ref{fig_emax_vs_a_cha}, \ref{fig_emax_vs_a_hau} and \ref{fig_emax_vs_a_qua} show the eccentricities $e_{\rm max}$ raised by 1$^{\rm st}$- and 2$^{\rm nd}$-order resonances around Chariklo, Haumea and Quaoar.
They are the functions shown in Figs~\ref{fig_DJ_emax_1st_order} and
\ref{fig_DJ_emax_2nd_order}, relevant for each resonance.
The resonant radii are calculated using the quadrupole gravitational potential in the ellipsoid case (see expression~\ref{eq_pot_ell}) and the potential $-GM/r$ in the mass anomaly case, together with the condition given by Eq.~\ref{eq_SOR}.
Based on the values listed in Table~\ref{tab_param_cha_hau_qua}, we also plot in Figs.~\ref{fig_emax_vs_a_cha}, \ref{fig_emax_vs_a_hau} and \ref{fig_emax_vs_a_qua} the radii of the rings observed around the three bodies together with the nearby resonances 1/3, and in the case of Quaoar, the location of the 5/7 SOR resonance that lies close to the ring Q2R.

\subsection{Chariklo}

Figure~\ref{fig_emax_vs_a_cha} displays trajectories of corotating particles encircling the fixed points $C_2$ and $C_4$. From Table~\ref{tab_param_cha_hau_qua}, we obtain $q^{2/3} C_{22} \sim 0.007$. This value can be used to assess the dynamical stability of the corotation points $C_2$ and $C_4$, as expressed by the condition~\ref{eq_C22_critic}. Since it is not met (by a small margin), the points are $C_2$ and $C_4$ are expected to be unstable. More accurate observations are needed to pin down the value of $q^{2/3} C_{22}$, and thus assess more precisely the dynamical stability of Chariklo's corotation points.
We note that even if $C_2$ and $C_4$ are dynamically stable, they correspond in any case to local maxima of potential energy. As such, they are expected to be unstable against the dissipative effect of collisions, a conclusion that also holds for Haumea and Quaoar.

The figure~\ref{fig_emax_vs_a_cha} reveals a dense mesh of 1$^{\rm st}$-order SORs bracketing the synchronous orbit. As discussed in \cite{sicardy2019}, these resonances cause torques that rapidly clear the corotation zone, pushing material towards Chariklo inside the synchronous orbit and repelling that material towards outer regions outside the synchronous orbit. 
The clearing time scales are a few tens of years for resonances associated 
with Chariklo's triaxiality, and a few million years for resonances associated 
with the mass anomaly of $\mu= 5 \times 10^{-3}$ considered in 
\cite{sicardy2019}\footnote{Other clearing time scales are obtained knowing they scale like $\mu^{-2}$, since the torques at Lindblad resonances scale like $\mu^2$.}.

Moving outwards, we see that the 2$^{\rm nd}$-order 2/4 resonance associated with Chariklo's triaxiality near the orbital radius 310~km induces large eccentricities of more than 0.2 on the particles, which prevents the presence of stable ring in this zone.

A more quiescent situation then sets in beyond the 2/4 resonance region.  
The 2$^{\rm nd}$-order 1/3 resonance associated with a mass anomaly is then the only 
remaining one found in the pool of 1$^{\rm st}$-order or 2$^{\rm nd}$-order resonances.
With $\mu= 10^{-3}$, it excites a moderate eccentricity of 0.01.
As discussed earlier, the fourth-order 2/6 resonance associated with Chariklo's triaxiality has a negligible effect on a ring in spite of a large value of $\epsilon_{\rm elon}$ in Eq.~\ref{eq_H_Theta_phi}. This is confirmed by $N$-body simulations of Paper~II.
Conversely, the simulations show the 1/3 resonance have a confining effect on
a collisional disk, in spite of the expected streamline crossing problem. 

We note that at the moment, the uncertainty on the 1/3 resonance location (the purple region in Fig.~\ref{fig_emax_vs_a_cha}) is consistent with Chariklo's rings being trapped at the 1/3 resonance. A more accurate determination of the resonance location, deduced from a more accurate determination of Chariklo's mass, is now needed to confirm this point.

\subsection{Haumea}

Figure~\ref{fig_emax_vs_a_hau} is the equivalent of Fig.~\ref{fig_emax_vs_a_cha} for Haumea. We now have $q^{2/3} C_{22} \sim 0.026$ (Table~\ref{tab_param_cha_hau_qua}), so that Haumea's corotation points $C_2$ and $C_4$ are unstable by a large margin from Eq.~\ref{eq_C22_critic}. This makes the entire corotation region of Haumea inappropriate for hosting rings.
Moreover, Fig.~\ref{fig_emax_vs_a_hau} shows that the 2$^{\rm nd}$-order resonance 2/4 raises eccentricities as high as 0.6. This makes the all region inside the radius $\sim$2050~km inhospitable for rings. As in the case of Chariklo, the 1/3 SOR associated with a Haumea mass anomaly is the only one that induces moderate eccentricities (here of the order of 0.01), which may explain why a ring can be observed near that resonance.

\begin{figure}
\centerline{\includegraphics[totalheight=58mm,trim=0 0 0 0]{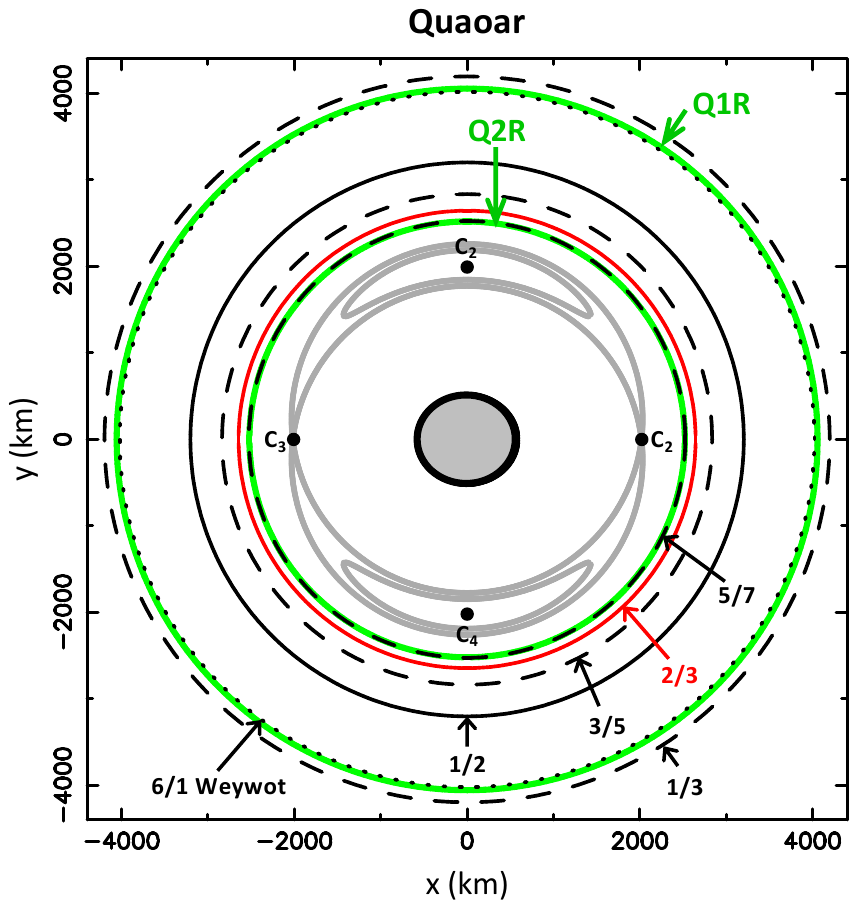}}
\centerline{\includegraphics[totalheight=55mm,trim=0 0 0 0]{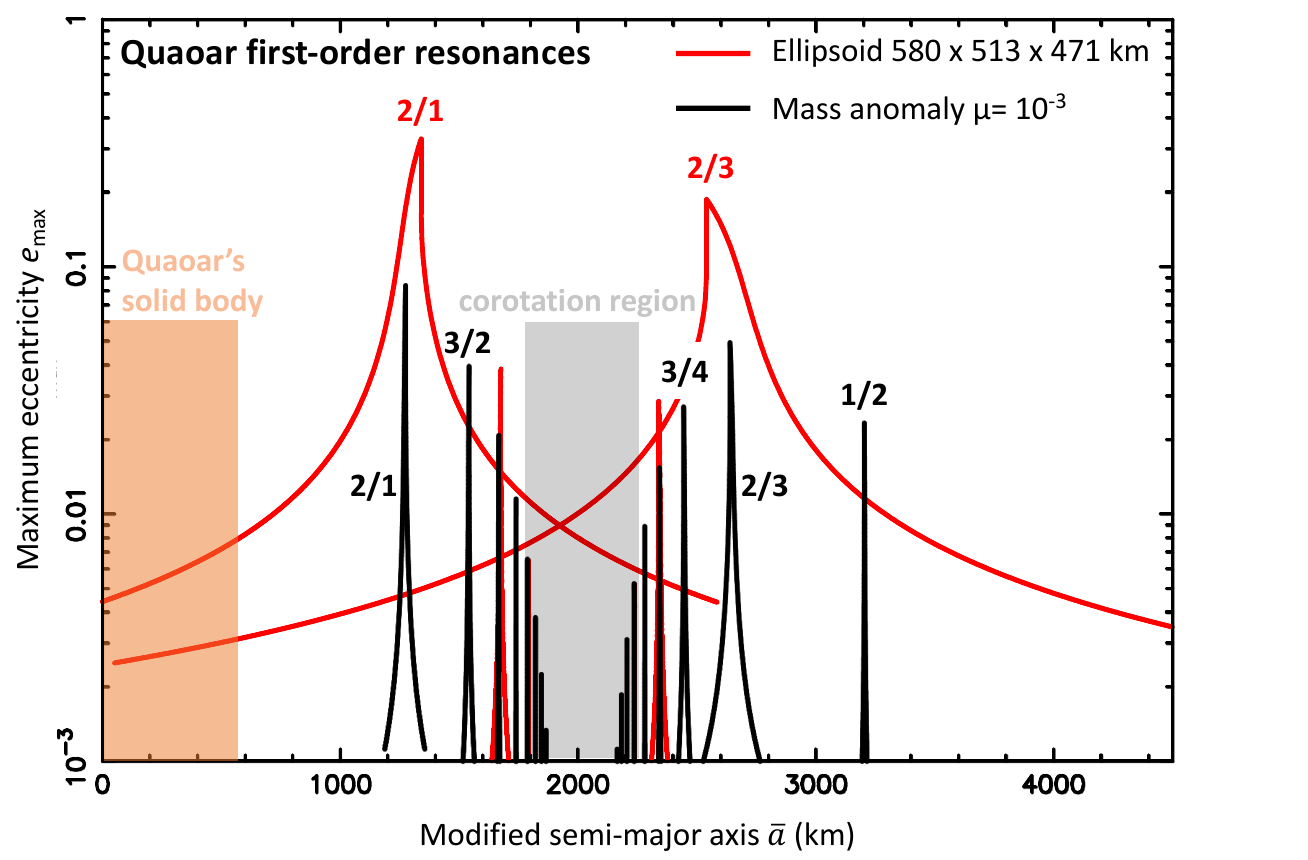}}
\centerline{\includegraphics[totalheight=55mm,trim=0 0 0 0]{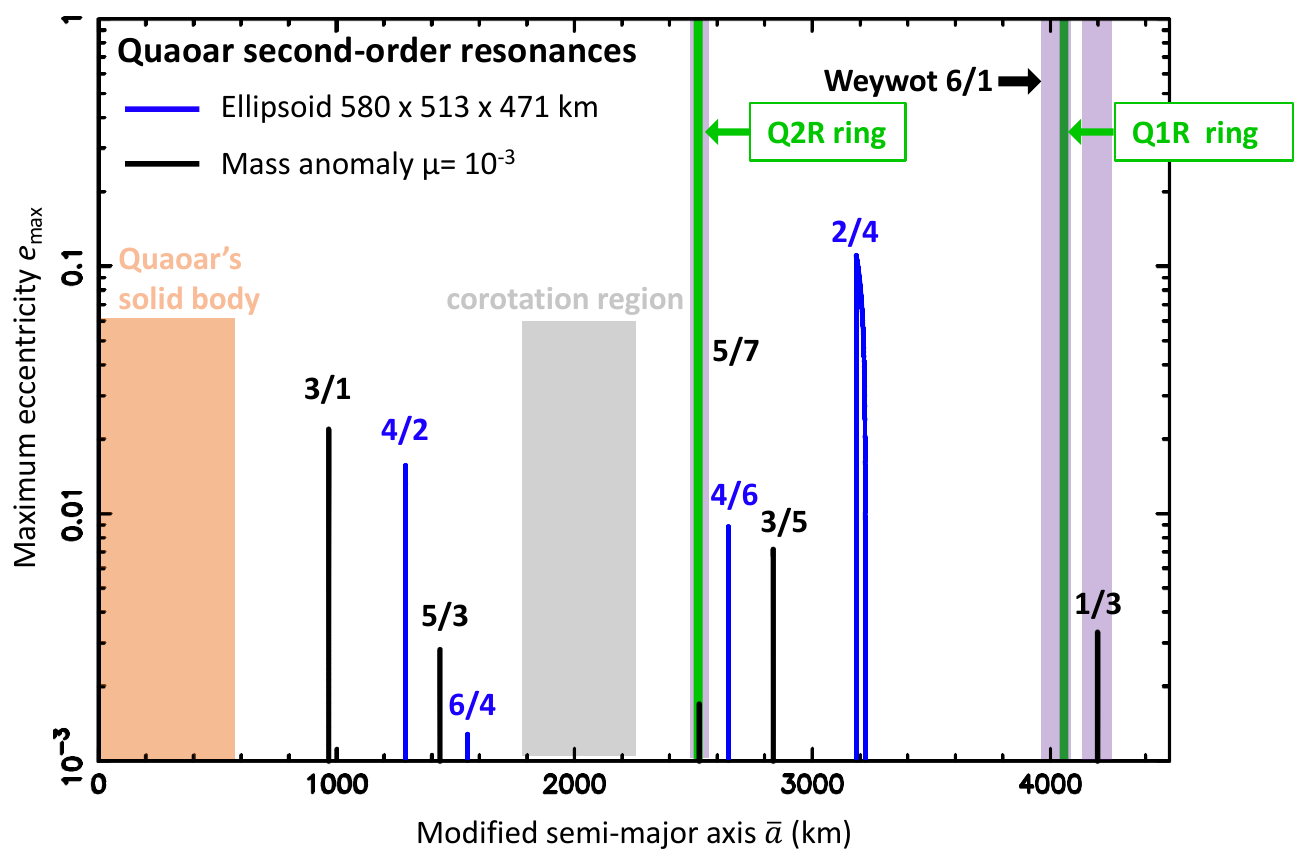}}
\caption{
The same as Fig.~\ref{fig_emax_vs_a_cha} for Quaoar, where two resonances have been added in the upper panel:
the fifth-order resonance 6/1 with Weywot (the dotted circle close to Q1R), and
the second-order resonance 5/7 with Quaoar, the dashed circle close to Q2R.
As for Chariklo, the uncertainties on the 5/7 and 1/3 resonance radii (purple boxes in the lower panel) are larger than the uncertainties on the Q1R and Q2R ring radii, see Table~\ref{tab_param_cha_hau_qua}.
}
\label{fig_emax_vs_a_qua}
\end{figure}

\subsection{Quaoar}

Table~\ref{tab_param_cha_hau_qua} now provides $q^{2/3} C_{22} \sim 0.0009$, so that from Eq.~\ref{eq_C22_critic} the Quaoar's corotation points $C_2$ and $C_4$ shown in Fig.~\ref{fig_emax_vs_a_qua} appear to be safely stable.
Concerning the SORs, the fact that Quaoar is a slower rotator than Chariklo and Haumea places these resonances farther out, when compared to the radius of the body. Consequently, Quaoar has a generally quieter environment than those of Chariklo and Haumea, due to a less dense mesh of resonances. However, the 1$^{\rm st}$- and 2$^{\rm nd}$-order resonances associated with Quaoar's triaxiality still excite large eccentricities, as do the 1$^{\rm st}$-order SORs associated with a putative Quaoar mass anomaly of $\mu= 10^{-3}$.

We see in Fig.~\ref{fig_emax_vs_a_qua} that Quaoar's rings Q1R and Q2R are both close to 2$^{\rm nd}$-order SORs (1/3 and 5/7, respectively). Both resonances excite modest eccentricities well below 0.01, assuming $\mu= 10^{-3}$.
We note that although Q1R is formally outside the purple region defining the possible radial location of the 1/3 SOR, the mismatch is at the 2.4$\sigma$ level when accounting for the error bars, so it remains marginally significant.

The ring Q1R is also coincident (at the 1$\sigma$ level) with the inner fifth-order 6/1 Mean Motion Resonance (MMR) with the satellite Weywot (Fig.~\ref{fig_emax_vs_a_qua}). 
The respective effects of the Weywot 6/1 MMR and the 1/3 Quaoar's SOR depend on Weywot's orbital eccentricity $e_{\rm W}$ and on Quaoar's mass anomaly $\mu$, respectively \citep{morgado2023}. 
Concerning Weywot, only an upper limit $e_{\rm W} \lesssim 0.034$ is currently available \citep{braga2025}.
With $e_{\rm W} \sim 0.005$ and $\mu \sim 10^{-3}$, the 6/1 MMR and the 1/3 SOR excite comparable eccentricities on ring particles, so that none of them can a priory be neglected compared to the other \citep{morgado2023}. 
More detailed calculations by \cite{rodriguez2023} show that in case of an orbital eccentricity of Weywot, a multiplet of six MMRs appears near the orbit of Q1R, possibly causing a clumping of particles in arcs. These effects are not considered in this paper, as we restrict our analysis to 1$^{\rm st}$ and 2$^{\rm nd}$ resonances.

Concerning Q2R, its radius coincides (below the 1$\sigma$ level) with the location of the 2$^{\rm nd}$-order 5/7 SOR. This resonance is actually bracketed by the stronger 1$^{\rm st}$-order resonances 3/4 and 2/3, a topic numerically discussed in Paper~II.

\section{Conclusions}
\label{sec_conclusions}

This paper investigates from an analytical standpoint the behavior of test particles near $j^{\rm th}$-order Spin-Orbit Resonances (SORs) between a non-axisymmetric body and test particles.

We have estimated the stability of the corotation Lagrange points associated with 
the triaxial shape of Chariklo, Haumea and Quaoar.
Chariklo's Lagrange points $C_2$ and $C_4$ (Fig.~\ref{fig_emax_vs_a_cha}) are marginally unstable adopting the current knowledge of its shape. However, this may change as updated shape models are obtained.
Conversely, Haumea's Lagrange points (Fig.~\ref{fig_emax_vs_a_hau}) are highly unstable, making the entire corotation region inhospitable for ring material.
Finally Quaoar's Lagrange points (Fig.~\ref{fig_emax_vs_a_qua}) are dynamically stable and could support the presence of moonlets librating around the $C_2$ and $C_4$ points.
However, in all these cases, the $C_2$ and $C_4$ points correspond to maxima of potential, and are in principle unstable under the effect of dissipative collisions, unless energy is supplied by hypothetical moonlets as may be the case for Neptune's ring arcs \citep{renner2004,renner2014,depater2018}.

We have examined the topology of phase portraits for SORs of orders ranging from $j=1$ to $j=5$, the cases $j>5$ being a mere repetition of what is observed for $j=5$ (Fig.~\ref{fig_maps_hamiltonians}).
This examination shows that only 1$^{\rm st}$-order Lindblad ($j=1$) and 2$^{\rm nd}$-order ($j=2$) SORs can excite initially circular orbits. As such, they are the only ones that are expected to significantly disturb a dense collisional ring.
In this context, we have estimated the characteristic widths as well as the typical eccentricities excited at these resonances, see Figs.~\ref{fig_DJ_emax_1st_order} and \ref{fig_DJ_emax_2nd_order}, and Table~\ref{tab_resonance_W_emax}.

Applications to Chariklo, Haumea and Quaoar are made. 
Concerning Chariklo and Haumea, the mesh of 1$^{\rm st}$-order and 2$^{\rm nd}$-order SORs is dense, 
where the strong 1$^{\rm st}$-order resonances excite high orbital eccentricities.
For these two bodies, this makes the region inside of the 1/2 (2/4 in the case of a triaxial body) SOR a strongly perturbed zone (Figs.~\ref{fig_emax_vs_a_cha} and \ref{fig_emax_vs_a_hau}). In that context, the 2$^{\rm nd}$-order 1/3 resonance is the only one that does not excite high eccentricities, being at the same time separated from the perturbed region.

In the case of Quaoar, SORs are more widely separated, making its entire surrounding a quieter place compared to Chariklo and Haumea. Even though, the 2/1 and 2/3, 1/2 and 2/4 SORs excite large eccentricities (Fig.~\ref{fig_emax_vs_a_qua}) that should strongly perturb a ring. Conversely, the 5/7 and 1/3 2$^{\rm nd}$-order SORs (near the Q2R and Q1R rings, respectively) have a less drastic effect.

We show that unlike to 1$^{\rm st}$-order resonances, the periodic orbits corresponding to 2$^{\rm nd}$-order resonances have a self-crossing point (Fig.~\ref{fig_periodic_orbits}). This issue is explored numerically in Paper~II using $N$-body collisional simulations. In particular, we will show that in spite of the self-crossing problem, ring confinement is in fact possible near the 1/3 resonance. 

\begin{acknowledgements}
This work has been supported by the French ANR project Roche, number ANR-23-CE49-0012.
\end{acknowledgements}

%\bibliographystyle{aa}
%\bibliography{references}

\begin{thebibliography}{33}
\expandafter\ifx\csname natexlab\endcsname\relax\def\natexlab#1{#1}\fi

\bibitem[{{Balmino}(1994)}]{balmino1994}
{Balmino}, G. 1994, Celestial Mechanics and Dynamical Astronomy, 60, 331

\bibitem[{{Boyce}(1997)}]{boyce1997}
{Boyce}, W. 1997, Celestial Mechanics and Dynamical Astronomy, 67, 107

\bibitem[{{Braga-Ribas} {et~al.}(2014){Braga-Ribas}, {Sicardy}, {Ortiz},
  {Snodgrass}, {Roques}, {Vieira-Martins}, {Camargo}, {Assafin}, {Duffard},
  {Jehin}, {Pollock}, {Leiva}, {Emilio}, {Machado}, {Colazo}, {Lellouch},
  {Skottfelt}, {Gillon}, {Ligier}, {Maquet}, {Benedetti-Rossi}, {Gomes},
  {Kervella}, {Monteiro}, {Sfair}, {El Moutamid}, {Tancredi}, {Spagnotto},
  {Maury}, {Morales}, {Gil-Hutton}, {Roland}, {Ceretta}, {Gu}, {Wang},
  {Harps{\o}e}, {Rabus}, {Manfroid}, {Opitom}, {Vanzi}, {Mehret}, {Lorenzini},
  {Schneiter}, {Melia}, {Lecacheux}, {Colas}, {Vachier}, {Widemann},
  {Almenares}, {Sandness}, {Char}, {Perez}, {Lemos}, {Martinez},
  {J{\o}rgensen}, {Dominik}, {Roig}, {Reichart}, {Lacluyze}, {Haislip},
  {Ivarsen}, {Moore}, {Frank}, \& {Lambas}}]{braga2014}
{Braga-Ribas}, F., {Sicardy}, B., {Ortiz}, J.~L., {et~al.} 2014, \nat, 508, 72

\bibitem[{{Braga-Ribas} {et~al.}(2025){Braga-Ribas}, {Vachier}, {Desmars},
  {Margoti}, \& {Sicardy}}]{braga2025}
{Braga-Ribas}, F., {Vachier}, F., {Desmars}, J., {Margoti}, G., \& {Sicardy},
  B. 2025, Philosophical Transactions of the Royal Society of London Series A,
  383, 20240200

\bibitem[{{De Pater} {et~al.}(2018){De Pater}, {Renner}, {Showalter}, \&
  {Sicardy}}]{depater2018}
{De Pater}, I., {Renner}, S., {Showalter}, M.~R., \& {Sicardy}, B. 2018, in
  Planetary Ring Systems. Properties, Structure, and Evolution, ed. M.~S.
  {Tiscareno} \& C.~D. {Murray} (Cambridge University Press), 112--124

\bibitem[{{Dermott} \& {Murray}(1981)}]{dermott1981}
{Dermott}, S.~F. \& {Murray}, C.~D. 1981, \icarus, 48, 1

\bibitem[{{El Moutamid} {et~al.}(2014){El Moutamid}, {Sicardy}, \&
  {Renner}}]{elmoutamid2014}
{El Moutamid}, M., {Sicardy}, B., \& {Renner}, S. 2014, Celestial Mechanics and
  Dynamical Astronomy, 118, 235

\bibitem[{{Ellis} \& {Murray}(2000)}]{ellis2000}
{Ellis}, K.~M. \& {Murray}, C.~D. 2000, \icarus, 147, 129

\bibitem[{{Ferraz-Mello}(1985)}]{ferraz1985}
{Ferraz-Mello}, S. 1985, Celestial Mechanics, 35, 209

\bibitem[{{Leiva} {et~al.}(2017){Leiva}, {Sicardy}, {Camargo}, {Ortiz},
  {Desmars}, {B{\'e}rard}, {Lellouch}, {Meza}, {Kervella}, {Snodgrass},
  {Duffard}, {Morales}, {Gomes-J{\'u}nior}, {Benedetti-Rossi},
  {Vieira-Martins}, {Braga-Ribas}, {Assafin}, {Morgado}, {Colas}, {De Witt},
  {Sickafoose}, {Breytenbach}, {Dauvergne}, {Schoenau}, {Maquet}, {Bath},
  {Bode}, {Cool}, {Lade}, {Kerr}, \& {Herald}}]{leiva2017}
{Leiva}, R., {Sicardy}, B., {Camargo}, J.~I.~B., {et~al.} 2017, \aj, 154, 159

\bibitem[{{Lemaitre}(1984)}]{lemaitre1984}
{Lemaitre}, A. 1984, Celestial Mechanics, 32, 109

\bibitem[{{Morgado} {et~al.}(2021){Morgado}, {Sicardy}, {Braga-Ribas},
  {Desmars}, {Gomes-J{\'u}nior}, {B{\'e}rard}, {Leiva}, {Ortiz},
  {Vieira-Martins}, {Benedetti-Rossi}, {Santos-Sanz}, {Camargo}, {Duffard},
  {Rommel}, {Assafin}, {Boufleur}, {Colas}, {Kretlow}, {Beisker}, {Sfair},
  {Snodgrass}, {Morales}, {Fern{\'a}ndez-Valenzuela}, {Amaral}, {Amarante},
  {Artola}, {Backes}, {Bath}, {Bouley}, {Buie}, {Cacella}, {Colazo}, {Colque},
  {Dauvergne}, {Dominik}, {Emilio}, {Erickson}, {Evans}, {Fabrega-Polleri},
  {Garcia-Lambas}, {Giacchini}, {Hanna}, {Herald}, {Hesler}, {Hinse},
  {Jacques}, {Jehin}, {J{\o}rgensen}, {Kerr}, {Kouprianov}, {Levine}, {Linder},
  {Maley}, {Machado}, {Maquet}, {Maury}, {Melia}, {Meza}, {Mondon}, {Moura},
  {Newman}, {Payet}, {Pereira}, {Pollock}, {Poltronieri}, {Quispe-Huaynasi},
  {Reichart}, {de Santana}, {Schneiter}, {Sieyra}, {Skottfelt}, {Soulier},
  {Starck}, {Thierry}, {Torres}, {Trabuco}, {Unda-Sanzana}, {Yamashita},
  {Winter}, {Zapata}, \& {Zuluaga}}]{morgado2021}
{Morgado}, B.~E., {Sicardy}, B., {Braga-Ribas}, F., {et~al.} 2021, \aap, 652,
  A141

\bibitem[{{Morgado} {et~al.}(2023){Morgado}, {Sicardy}, {Braga-Ribas}, {Ortiz},
  {Salo}, {Vachier}, {Desmars}, {Pereira}, {Santos-Sanz}, {Sfair}, {de
  Santana}, {Assafin}, {Vieira-Martins}, {Gomes-J{\'u}nior}, {Margoti},
  {Dhillon}, {Fern{\'a}ndez-Valenzuela}, {Broughton}, {Bradshaw}, {Langersek},
  {Benedetti-Rossi}, {Souami}, {Holler}, {Kretlow}, {Boufleur}, {Camargo},
  {Duffard}, {Beisker}, {Morales}, {Lecacheux}, {Rommel}, {Herald}, {Benz},
  {Jehin}, {Jankowsky}, {Marsh}, {Littlefair}, {Bruno}, {Pagano}, {Brandeker},
  {Collier-Cameron}, {Flor{\'e}n}, {Hara}, {Olofsson}, {Wilson}, {Benkhaldoun},
  {Busuttil}, {Burdanov}, {Ferrais}, {Gault}, {Gillon}, {Hanna}, {Kerr},
  {Kolb}, {Nosworthy}, {Sebastian}, {Snodgrass}, {Teng}, \& {de
  Wit}}]{morgado2023}
{Morgado}, B.~E., {Sicardy}, B., {Braga-Ribas}, F., {et~al.} 2023, \nat, 614,
  239

\bibitem[{{Murray} \& {Dermott}(2000)}]{murray2000}
{Murray}, C.~D. \& {Dermott}, S.~F. 2000, {Solar System Dynamics} (Cambridge
  University Press)

\bibitem[{{Ortiz} {et~al.}(2003){Ortiz}, {Guti{\'e}rrez}, {Sota}, {Casanova},
  \& {Teixeira}}]{ortiz2003}
{Ortiz}, J.~L., {Guti{\'e}rrez}, P.~J., {Sota}, A., {Casanova}, V., \&
  {Teixeira}, V.~R. 2003, \aap, 409, L13

\bibitem[{{Ortiz} {et~al.}(2023){Ortiz}, {Pereira}, {Sicardy}, {Braga-Ribas},
  {Takey}, {Fouad}, {Shaker}, {Kaspi}, {Brosch}, {Kretlow}, {Leiva}, {Desmars},
  {Morgado}, {Morales}, {Vara-Lubiano}, {Santos-Sanz},
  {Fern{\'a}ndez-Valenzuela}, {Souami}, {Duffard}, {Rommel}, {Kilic}, {Erece},
  {Koseoglu}, {Ege}, {Morales}, {Alvarez-Candal}, {Rizos},
  {G{\'o}mez-Lim{\'o}n}, {Assafin}, {Vieira-Martins}, {Gomes-J{\'u}nior},
  {Camargo}, \& {Lecacheux}}]{ortiz2023}
{Ortiz}, J.~L., {Pereira}, C.~L., {Sicardy}, B., {et~al.} 2023, \aap, 676, L12

\bibitem[{{Ortiz} {et~al.}(2017){Ortiz}, {Santos-Sanz}, {Sicardy},
  {Benedetti-Rossi}, {B{\'e}rard}, {Morales}, {Duffard}, {Braga-Ribas}, {Hopp},
  {Ries}, {Nascimbeni}, {Marzari}, {Granata}, {P{\'a}l}, {Kiss}, {Pribulla},
  {Kom{\v{z}}{\'\i}k}, {Hornoch}, {Pravec}, {Bacci}, {Maestripieri}, {Nerli},
  {Mazzei}, {Bachini}, {Martinelli}, {Succi}, {Ciabattari}, {Mikuz},
  {Carbognani}, {Gaehrken}, {Mottola}, {Hellmich}, {Rommel},
  {Fern{\'a}ndez-Valenzuela}, {Campo Bagatin}, {Cikota}, {Cikota}, {Lecacheux},
  {Vieira-Martins}, {Camargo}, {Assafin}, {Colas}, {Behrend}, {Desmars},
  {Meza}, {Alvarez-Candal}, {Beisker}, {Gomes-Junior}, {Morgado}, {Roques},
  {Vachier}, {Berthier}, {Mueller}, {Madiedo}, {Unsalan}, {Sonbas}, {Karaman},
  {Erece}, {Koseoglu}, {Ozisik}, {Kalkan}, {Guney}, {Niaei}, {Satir},
  {Yesilyaprak}, {Puskullu}, {Kabas}, {Demircan}, {Alikakos}, {Charmandaris},
  {Leto}, {Ohlert}, {Christille}, {Szak{\'a}ts}, {Tak{\'a}csn{\'e} Farkas},
  {Varga-Vereb{\'e}lyi}, {Marton}, {Marciniak}, {Bartczak}, {Santana-Ros},
  {Butkiewicz-B{\k{a}}k}, {Dudzi{\'n}ski}, {Al{\'\i}-Lagoa}, {Gazeas},
  {Tzouganatos}, {Paschalis}, {Tsamis}, {S{\'a}nchez-Lavega},
  {P{\'e}rez-Hoyos}, {Hueso}, {Guirado}, {Peris}, \&
  {Iglesias-Marzoa}}]{ortiz2017}
{Ortiz}, J.~L., {Santos-Sanz}, P., {Sicardy}, B., {et~al.} 2017, \nat, 550, 219

\bibitem[{{Pereira} {et~al.}(2023){Pereira}, {Sicardy}, {Morgado},
  {Braga-Ribas}, {Fern{\'a}ndez-Valenzuela}, {Souami}, {Holler}, {Boufleur},
  {Margoti}, {Assafin}, {Ortiz}, {Santos-Sanz}, {Epinat}, {Kervella},
  {Desmars}, {Vieira-Martins}, {Kilic}, {Gomes J{\'u}nior}, {Camargo},
  {Emilio}, {Vara-Lubiano}, {Kretlow}, {Albert}, {Alcock}, {Ball}, {Bender},
  {Buie}, {Butterfield}, {Camarca}, {Castro-Chac{\'o}n}, {Dunford}, {Fisher},
  {Gamble}, {Geary}, {Gnilka}, {Green}, {Hartman}, {Huang}, {Januszewski},
  {Johnston}, {Kagitani}, {Kamin}, {Kavelaars}, {Keller}, {de Kleer}, {Lehner},
  {Luken}, {Marchis}, {Marlin}, {McGregor}, {Nikitin}, {Nolthenius}, {Patrick},
  {Redfield}, {Rengstorf}, {Reyes-Ruiz}, {Seccull}, {Skrutskie}, {Smith},
  {Sproul}, {Stephens}, {Szentgyorgyi}, {S{\'a}nchez-Sanju{\'a}n}, {Tatsumi},
  {Verbiscer}, {Wang}, {Yoshida}, {Young}, \& {Zhang}}]{pereira2023}
{Pereira}, C.~L., {Sicardy}, B., {Morgado}, B.~E., {et~al.} 2023, \aap, 673, L4

\bibitem[{{Pfenniger}(1984)}]{pfenniger1984}
{Pfenniger}, D. 1984, \aap, 134, 373

\bibitem[{{Renner} \& {Sicardy}(2004)}]{renner2004}
{Renner}, S. \& {Sicardy}, B. 2004, Celestial Mechanics and Dynamical
  Astronomy, 88, 397

\bibitem[{{Renner} {et~al.}(2014){Renner}, {Sicardy}, {Souami}, {Carry}, \&
  {Dumas}}]{renner2014}
{Renner}, S., {Sicardy}, B., {Souami}, D., {Carry}, B., \& {Dumas}, C. 2014,
  \aap, 563, A133

\bibitem[{{Rodr{\'\i}guez} {et~al.}(2023){Rodr{\'\i}guez}, {Morgado}, \&
  {Callegari}}]{rodriguez2023}
{Rodr{\'\i}guez}, A., {Morgado}, B.~E., \& {Callegari}, Jr., N. 2023, \mnras,
  525, 3376

\bibitem[{{Salo} \& {Sicardy}(2024)}]{salo2024}
{Salo}, H. \& {Sicardy}, B. 2024, in European Planetary Science Congress,
  EPSC2024--534

\bibitem[{{Salo} \& {Sicardy}(2025)}]{salo2025}
{Salo}, H. \& {Sicardy}, B. 2025, \aap, submitted (paper II)

\bibitem[{{Salo} {et~al.}(2021){Salo}, {Sicardy}, {Mondino-Llermanos}, {Soumi},
  {Renner}, {Morgado}, {Braga-Ribas}, {Benedetti-Rossi}, \& {de
  Santana}}]{salo2021}
{Salo}, H., {Sicardy}, B., {Mondino-Llermanos}, A., {et~al.} 2021, in European
  Planetary Science Congress, EPSC2021--338

\bibitem[{{Sicardy}(2020)}]{sicardy2020a}
{Sicardy}, B. 2020, \aj, 159, 102

\bibitem[{{Sicardy} {et~al.}(2024){Sicardy}, {Braga-Ribas}, {Buie}, {Ortiz}, \&
  {Roques}}]{sicardy2024b}
{Sicardy}, B., {Braga-Ribas}, F., {Buie}, M.~W., {Ortiz}, J.~L., \& {Roques},
  F. 2024, \aapr, 32, 6

\bibitem[{{Sicardy} {et~al.}(2018){Sicardy}, {El Moutamid}, {Quillen},
  {Schenk}, {Showalter}, \& {Walsh}}]{sicardy2018}
{Sicardy}, B., {El Moutamid}, M., {Quillen}, A.~C., {et~al.} 2018, in Planetary
  Ring Systems. Properties, Structure, and Evolution, ed. M.~S. {Tiscareno} \&
  C.~D. {Murray} (Cambridge University Press), 135--154

\bibitem[{{Sicardy} {et~al.}(2019){Sicardy}, {Leiva}, {Renner}, {Roques}, {El
  Moutamid}, {Santos-Sanz}, \& {Desmars}}]{sicardy2019}
{Sicardy}, B., {Leiva}, R., {Renner}, S., {et~al.} 2019, Nature Astronomy, 3,
  146

\bibitem[{{Sicardy} {et~al.}(2020){Sicardy}, {Renner}, {Leiva}, {Roques}, {El
  Moutamid}, {Santos-Sanz}, \& {Desmars}}]{sicardy2020b}
{Sicardy}, B., {Renner}, S., {Leiva}, R., {et~al.} 2020, in The Trans-Neptunian
  Solar System, ed. D.~{Prialnik}, M.~A. {Barucci}, \& L.~{Young} (Elsevier),
  249--269

\bibitem[{{Sicardy} \& {Salo}(2024)}]{sicardy2024a}
{Sicardy}, B. \& {Salo}, H. 2024, in European Planetary Science Congress,
  EPSC2024--123

\bibitem[{{Sicardy} {et~al.}(2021){Sicardy}, {Salo}, {Souami}, {Renner},
  {Morgado}, {Braga-Ribas}, {Benedetti-Rossi}, \& {de Santana}}]{sicardy2021}
{Sicardy}, B., {Salo}, H., {Souami}, D., {et~al.} 2021, in European Planetary
  Science Congress, EPSC2021--91

\bibitem[{{Vachier} {et~al.}(2012){Vachier}, {Berthier}, \&
  {Marchis}}]{vachier2012}
{Vachier}, F., {Berthier}, J., \& {Marchis}, F. 2012, \aap, 543, A68

\end{thebibliography}

\begin{appendix}

\section{Potential caused by a mass anomaly}
\label{app_potential_mass_anomaly}

A mass anomaly $\mu$, or ``mascon", introduces a dipole term with lowest values $|m|=1$ in Eq.~\ref{eq_U0}. This mass anomaly can reside at the surface of the object or be embedded in the body. To give a simple physical interpretation, it can be seen as a hemispheric mountain of height $h>0$ or a depression of depth $h<0$ located in the body equatorial plane at a characteristic distance $R_{\rm ref}$ (Eq.~\ref{eq_R_ref}) from the body center. Its mass relative to the body is then (noting that $\mu$ can be positive or negative),
\begin{equation}
\mu \sim \frac{1}{2} \left( \frac{h}{R_{\rm ref}} \right)^3.
\label{eq_height_mountain}
\end{equation}
This mass anomaly revolves at the spin rate of the body, $\Omega_{\rm B}$, so that 
\citep{sicardy2019,sicardy2020a,sicardy2020b}
\begin{equation}
\begin{array}{l}
U(\textbf{r}) =  \displaystyle -\frac{GM}{r} \\ \\
 \displaystyle - \frac{GM\mu}{R_{\rm ref}} 
\left\{\frac{1}{2} \sum_{m=-\infty}^{+\infty} 
\left[b_{1/2}^{(m)}\left(\frac{r}{R_{\rm ref}}\right) - q \delta_{(|m|,1)}\left(\frac{r}{R_{\rm ref}}\right) \right] 
\cos (m\theta)\right\}, \\
\end{array}
\label{eq_pot_ma}
\end{equation}
where $b_{1/2}^{(m)}$ is the classical Laplace coefficient, $\delta_{(|m|,1)}$ is the Kronecker delta function that is associated with the indirect part of the potential and
\begin{equation}
q= \frac{\Omega_{\rm B}^2 R_{\rm ref}^3}{GM}
\label{eq_q}
\end{equation}
is the rotational parameter.

The expression to be used in Eq.~\ref{eq_Umj_gen} is then
\begin{equation}
U_{m}(\alpha) =  -\left(\frac{GM\mu}{2 R_{\rm ref}}\right) [b_{1/2}^{(m)}(\alpha) - q \delta_{(|m|,1)}\alpha].
\label{eq_U_m_alpha_ma}
\end{equation}

\section{Potential of a triaxial homogeneous ellipsoid}
\label{app_potential_ellipsoid}

We consider a homogeneous ellipsoid with principal semi-axes $A$, $B$ and $C$.
The elongation $C_{22}$ and the dynamical oblateness $J_2$ of the ellipsoid
are given by \cite{balmino1994},
$$
\begin{array}{ll}
\displaystyle C_{22}= \frac{A^2-B^2}{20 R_{\rm ref}^2} {\rm~~~and} &
\displaystyle J_2 = -C_{20} = \frac{A^2+B^2-2C^2}{10 R_{\rm ref}^2},
\end{array}
$$
where the reference radius $R_{\rm ref}$ is defined by
\begin{equation}
\frac{3}{R_{\rm ref}^2} = \frac{1}{A^2} + \frac{1}{B^2} + \frac{1}{C^2}.
\label{eq_R_ref}
\end{equation}
Following \cite{boyce1997}, \cite{sicardy2019}, \cite{sicardy2020a} and \cite{sicardy2020b} used the non-standard parameters $\epsilon_{\rm elon}= (A^2-B^2)/2R_{\rm ref}^2= 10C_{22}$ and $f= (A^2+B^2-2C^2)/4R_{\rm ref}^2= (5/2)J_2$ to characterize the elongation and the oblateness of the object, respectively. This avoided carrying the factors 10 and $5/2$ in the various expressions of the potential. Another advantage of $\epsilon_{\rm elon}$ and $f$ was that they have simple physical interpretations when they approach zero, $\epsilon_{\rm elon} \sim (A-B)/A$ and $f= (A-C)/A$.

Here we use the more standard parameters $C_{22}$ and $J_2$ to be in line with other works
published in the literature, especially when obtained during flybys by space missions.

The quadrupole gravitational potential $U({\bf r})$ of the ellipsoid is derived from \cite{balmino1994}
and \citep{boyce1997}, see also \cite{sicardy2019}, \cite{sicardy2020a} and \cite{sicardy2020b}. 
To zeroth order in $J_2$ (except for $m=0$, see below), it reads
\begin{equation}
U(\textbf{r}) =  
-\left( \frac{GM}{R_{\rm ref}} \right) \sum_{-\infty}^{+\infty} 
\left( \frac{R_{\rm ref}}{r} \right)^{|m|+1} 
S_{|m/2|} (10C_{22})^{|m/2|}
\cos\left(m \theta \right)
\label{eq_pot_ell}
\end{equation}
where only even values of $m$ are allowed due to the $\pi$-symmetry of the ellipsoid.
The factor $S_{p}$ is recursively calculated through
$$
S_{p+1} = 2 \frac{(p+1/4)(p+3/4)}{(p+1)(p+5/2)} \times S_{p}, {\rm~~with~~} S_0 = 1.
$$

The expression~\ref{eq_pot_ell} shows from Eq.~\ref{eq_U0} that the term $U_m(\alpha)$ 
to be used in Eq.~\ref{eq_Umj_gen} is now
\begin{equation}
U_m(\alpha) = -\left(\frac{GM}{R_{\rm ref}}\right) \frac{(10C_{22})^{|m|/2} S_{|m/2|}}{\alpha^{|m|+1}}.
\label{eq_U_m_alpha_ell}
\end{equation}

The axisymmetric part of the potential, corresponding to $m=0$, 
is given in \cite{sicardy2019} to any order in $f$. 
Keeping only the first-order term, we have
$$
U_0 (r)= -\frac{GM}{r} 
\left[1 + \frac{J_2}{2} \left(\frac{R_{\rm ref}}{r} \right)^2 \right],
$$
from which the mean motion and epicyclic frequencies
\begin{equation}
\begin{array}{l}
\displaystyle
n^2(r)= \frac{1}{r} \frac{dU_0(r)}{dr} =
\frac{GM}{r^3} \left[1 + \frac{3J_2}{2} \left(\frac{R_{\rm ref}}{r}\right)^2\right] \\
 \\
\displaystyle
\kappa^2(r)= \frac{1}{r^3} \frac{d(r^4 n^2)}{dr} =
\frac{GM}{r^3} \left[1 - \frac{3J_2}{2} \left(\frac{R_{\rm ref}}{r}\right)^2\right] \\
\\
\end{array}
\label{eq_n_kappa}
\end{equation}
are derived. 
From these expressions, the location of the SOR resonances can be calculated using Eq.~\ref{eq_SOR}.

\section{Strengths of resonances}
\label{app_strength_resonances}

The strength of a resonance is quantified by the coefficient $\epsilon$ defined in Eq.~\ref{eq_epsilon}. It is obtained using Eq.~\ref{eq_Umj_gen}, where the operators $F_N$ are given in Table~\ref{tab_Fn}. They are listed according to the labels $N$ used by \cite{murray2000} and \cite{ellis2000}.
These operators contain both multiplicative factors and 
the derivative operators $D=d/d\alpha, D^2=d^2/d\alpha^2,...,D^j=d^j/d\alpha^j$ for a given $j$,
see Table~\ref{tab_Fn}.

\begin{table}[!h]
\caption{
Operators $F_N$.
\label{tab_Fn}
}
\begin{tabular}{lll}
\hline \hline
Order & Resonant angle & $F_N$ \\
\hline
$j=$1 & $m\lambda' - (m-1)\lambda - \varpi$     & $F_{27}= (1/2)[-2m - \alpha D]$ \\
$j=$2 & $(m\lambda' - (m-2)\lambda - \varpi)/2$ & $F_{45}= (1/8)[-5m + 4m^2 + $ \\
  &                                         & $(-2 + 4m)\alpha D + \alpha^2 D^2]$ \\
\hline
\end{tabular}
\end{table}

In the case of a homogeneous triaxial ellipsoid, 
the derivatives $\alpha^pD^p$ reduce to multiplicative factors
because $U_m(\alpha)$ depends only on powers of $\alpha$ (Eq.~\ref{eq_U_m_alpha_ell}), so that.
$\alpha^pD^p= (-1)^p(|m|+1)...(|m|+p)$.

In the case of a mass anomaly, two terms appear in Eq.~\ref{eq_U_m_alpha_ma}: 
the Laplace coefficients $b_{1/2}^{(m)}(\alpha)$ and the indirect term proportional to $\alpha$.
Thus, for the indirect term, all the derivatives $\alpha^pD^p$ for $p \geq 2$ vanish. 
To obtain the derivative of the Laplace coefficients, we use the following 
recursive relations for $p \geq 1$:
\begin{equation}
\begin{array}{ll}
D^p b_\gamma^{(m)}= & \\
\gamma
\left[ D^{p-1} b_{\gamma+1}^{(m-1)} + D^{p-1} b_{\gamma+1}^{(m+1)} -2\alpha D^{p-1} b_{\gamma+1}^{(m)} 
- 2(p-1) D^{p-2} b_{\gamma+1}^{(m)} \right], & \\
\end{array}
\end{equation}
with the convention that $D^{0}= 1$.

\section{The Hamiltonian approach }
\label{app_hamiltonians}

Here we summarize and complement calculations made elsewhere, 
see e.g.  \cite{lemaitre1984}, \cite{ferraz1985}, \cite{murray2000} and \cite{elmoutamid2014}. 
The Hamiltonian describing the motion of the particle near a $m/(m-j)$ resonance is
$$
{\cal H}= -\frac{(GM)^2}{2\Lambda^2} + 
\overline{U}_{m,j}(\alpha) \left( \frac{2\Gamma}{\Lambda} \right)^{j/2} \cos(j \phi) - \dot{\varpi}_{\rm sec} \Gamma,
$$
where the resonant argument $\phi$ (which depends on $m$ and $j$) is given by Eq.~\ref{eq_phi}, and where $\Lambda = \sqrt{GMa}$ and $\Gamma= \sqrt{GMa}(1 -\sqrt{1-e^2})$. 
The pairs of conjugate variables of this Hamiltonian are then
$$
\begin{array}{l}
\displaystyle 
\lambda  \longleftrightarrow J= \Lambda + \left( \frac{m-j}{j} \right) \Gamma \\ \\
\displaystyle
\phi \longleftrightarrow \Theta=  \Gamma. \\
\end{array}
$$
As ${\cal H}$ does not depends on $\lambda$,
$J$ is a constant of motion called the Jacobi constant.
The actions $\Lambda$ and $J$ can be expanded near their values $\Lambda_0$ and $J_0$ 
at exact resonance, where $m\Omega_{\rm B} - (m-j)n_0 - j\dot{\varpi}_{\rm sec}= 0$.
Dropping constant terms, the Hamiltonian reads
$$
{\cal H} = 
-\frac{3}{2a_0^2} \left[ J-J_0 - \left( \frac{m-j}{j} \right) \Theta \right]^2 +
\overline{U}_{m,j}(\alpha) \left( \frac{2\Theta}{\Lambda_0} \right)^{j/2} \cos(j \phi),
$$
where $\Lambda_0 = a_0^2 n_0$ and
$J - J_0 = (a_0^2 n_0/2) [ (\Delta a/a_0) + (m-j/j) e^2 ]$,
with $\Delta a = a - a_0$.
The first term in ${\cal H}$ merely represents the Keplerian motion 
(slightly shifted by the precession term $\dot{\varpi}_{\rm sec}$),
while the second term describes the perturbation induced by the resonance,
at the lowest order $j$ in eccentricity since $\Theta \propto e^2$.

The actions $\Lambda_0$, $J - J_0$ and $\Theta$ and the Hamiltonian can be normalized to $a_0^2 n_0$. 
Adopting $\tau= n_0 t$ as a new time scale, we obtain 
a one-degree of freedom Hamiltonian with new moment $\Theta= e^2/2$ and its conjugate angle $\phi$, parameterized by the normalized Jacobi constant $\Delta J$: 
\begin{equation}
\begin{array}{rl}
{\cal H} (\Theta,\phi) = &
\displaystyle 
-\frac{3}{2} \left[ \Delta J - \left( \frac{m-j}{j} \right) \Theta \right]^2 + \epsilon  (2\Theta)^{j/2} \cos(j \phi) \\ \\
 = &
 \displaystyle 
-\frac{3}{2} \left[ \Delta J - \left( \frac{m-j}{2j} \right) e^2 \right]^2 + \epsilon  e^j \cos(j \phi),
\end{array}
\label{eq_H_Theta_phi}
\end{equation}
where $\epsilon$ is given by Eq.~\ref{eq_epsilon} and
$\Delta J = (1/2) [\Delta a/a_0 + ((m-j)/j) e^2]$, see also Eq.~\ref{eq_definition_Delta_J}.
The parameter $\Delta J$ measures the distance of the orbits to exact resonance,
and the equations of motion are now
\begin{equation}
\left\{
\begin{array}{l}
\displaystyle 
\dot{\Theta} =  -\frac{\partial {\cal H}}{\partial  \phi} \\ \\
\displaystyle
\dot{\phi} =  +\frac{\partial {\cal H}}{\partial \Theta},
\end{array}
\right.
\end{equation}
where the dots denote the derivative with respect to $\tau$, not $t$.

The Hamiltonian may also be written in terms of the mixed variables
\begin{equation}
X= e \cos(\phi)  {\rm ~~and~~} Y= e \sin(\phi),
\label{eq_def_X_Y}
\end{equation}
that define the eccentricity vector
\begin{equation}
{\bf e}= (X,Y).
\label{eq_ecc_vector}
\end{equation}

\begin{table}[!t]
\caption{
The polynomials $P(X,Y)$ appearing in Eq.~\ref{eq_H_X_Y}.
\label{tab_pertub-part}
}
\begin{tabular}{ll}
\hline \hline
Resonance order $j$ & $\displaystyle P(X,Y)$ \\
\hline
1 & $X$ \\
2 & $X^2 - Y^2$ \\
3 & $X^3 - 3X Y^2$ \\
4 & $X^4 + Y^4 - 6X^2 Y^2$ \\
5 & $X^5 - 5X^3 Y^2 +10X Y^4$ \\
\hline
\end{tabular}
\end{table}

The last term of the Hamiltonian~\ref{eq_H_Theta_phi} now contains the factor $e^j \cos(j \phi)$. Using the classical expansion
$$
\displaystyle
\cos(j\phi) = \sum_{k=0}^{{\rm int}(j/2)} (-1)^k C_j^k \cos^{j-2k} (\phi) \sin^{2k} (\phi),
$$
where ${\rm int}(j/2)$ is the integer part of $j/2$ and $ C_j^k = j!/k!(j-k)!$, we obtain
\begin{equation}
{\cal H} (X,Y) = \displaystyle -\frac{3}{2} \left[ \Delta J - \left( \frac{m-j}{2j} \right) \left( X^2 + Y^2 \right) \right]^2 + \epsilon P(X,Y)
\label{eq_H_X_Y}
\end{equation}
where $P(X,Y)$ is a homogeneous polynomial of degree $j$,
$$
P(X,Y) = \sum_{k=0}^{{\rm int}(j/2)} (-1)^k C_j^k X^{j-2k} Y^{2k}.
$$
The expressions of $P(X,Y)$ are given in Table~\ref{tab_pertub-part} up to order $j=5$.
The equations of motion are now
\begin{equation}
\left\{
\begin{array}{l}
\displaystyle 
\dot{X} = -\frac{\partial {\cal H}}{\partial  Y} \\ \\
\displaystyle
\dot{Y} = +\frac{\partial {\cal H}}{\partial X}.
\end{array}
\right.
\label{eq_motion_X_Y}
\end{equation}

The phase portraits of resonances, i.e. the level curves of ${\cal H}$ (Eq.\ref{eq_H_X_Y}) are plotted in Fig.~\ref{fig_maps_hamiltonians} for orders $j$ ranging from one to five. 

\section{The cubic equation}
\label{app_cubic_equation}

We give the classical expressions of the real roots of cubic equations 
reduced to their so-called depressed version
\begin{equation}
z^3 + pz + q = 0,
\label{eq_depressed_cubic}
\end{equation}
where $(p, q) \in \mathbb{R}^2$.
The number of real solutions depends on the discriminant
\begin{equation}
\Delta = -(4p^3 + 27q^2).
\label{eq_discriminant}
\end{equation}
For $\Delta < 0$ the equation~\ref{eq_depressed_cubic} has a single real solution
\begin{equation}
z =
\left( \frac{-q - \sqrt{-\Delta/27}}{2}  \right)^{1/3} +
\left( \frac{-q + \sqrt{-\Delta/27}}{2}  \right)^{1/3}.
\label{eq_one_real_solution}
\end{equation}
Here and in all the paper, the cubic root of a real number is understood as the real root, 
i.e. discarding the two roots with imaginary parts.

For $\Delta \geq 0$ (which requires $p \leq 0$) the equation~\ref{eq_depressed_cubic} has three real solutions
\begin{equation}
z_k = 2 \sqrt{\frac{-p}{3}} \cos \left[ \frac{1}{3} \arccos \left( \frac{3q}{2p} \sqrt{\frac{-3}{p}} \right) + \frac{2k\pi}{3} \right],
\label{eq_three_real_solutions}
\end{equation}
with $k \in  \{0,1,2\}$.

\section{Potential near the corotating radius}
\label{app_potential_corotation}

\subsection{Corotation trajectories}

We consider the potential $V({\bf r})$ felt by a particle in the frame corotation 
with the body at angular velocity $\Omega_{\rm B}$, see \cite{sicardy2019,sicardy2020b}.
Near the corotation radius $a_{\rm cor}$, we have
\begin{equation}
V({\bf r}) = U({\bf r}) - \frac{\Omega_{\rm B}^2 r^2}{2} \sim 
-\Omega_{\rm B}^2 a_{\rm cor}^2 
\left[ \frac{3}{2} \left(\frac{\Delta r}{a_{\rm cor}}\right)^2 + f(\theta) \right].
\label{eq_rotating_pot}
\end{equation}
In the case of a mass anomaly we have
$$
f(\theta)=
q^{-1/6} \left(\frac{1}{\sqrt{q^{1/3} + q^{-1/3} - 2 \cos\theta}} - q^{1/2} \cos \theta\right) \mu
$$
and in the case of an ellipsoid we have 
\begin{equation}
f(\theta)=
2 \sum_{m=2}^{+\infty} q^{m/3} S_{m/2} (10 C_{22})^{m/2} \cos (m\theta) \sim
3 q^{2/3} C_{22} \cos(2\theta),
\label{eq_f_theta_ell}
\end{equation}
where only even values of $m$ are allowed. The approximation above is obtained by retaining only 
the lowest-order term $m=2$ (with $S_1=0.15$) in the summation, which is sufficient for order of magnitude considerations.

For typical values of $\mu$ and $C_{22}$, the corotation potential is largely dominated by $C_{22}$ \citep{sicardy2019}, so we use the expression~\ref{eq_f_theta_ell} for $f(\theta)$. Near $a_{\rm cor}$, and provided that the corotation point near the maximum of $V({\bf r})$ is dynamically stable, a particle with orbital semi-major axis $a$ follows a trajectory defined by
\begin{equation}
\frac{3}{8} \left(\frac{\Delta a}{a_{\rm cor}}\right)^2 + f(\theta) = {\rm constant},
\label{eq_trajec_corot}
\end{equation}
where $\Delta a = a - a_{\rm cor}$ \citep{dermott1981}.
In the ellipsoid case (Eq.~\ref{eq_f_theta_ell}), this implies a corotation region of full width
\begin{equation}
W_{\rm cor} \sim 8 R_{\rm ref} \sqrt{C_{22}}
\label{eq_full_width_corot}
\end{equation}
around the fixed points $C_2$ and $C_4$ displayed in Figs.~\ref{fig_emax_vs_a_cha}, \ref{fig_emax_vs_a_hau} and \ref{fig_emax_vs_a_qua}.

\subsection{Stability of corotation points}

The corotation points corresponding to local maxima of $V({\bf r})$ are linearly stable 
as long as \citep{murray2000}
\begin{equation}
\left(4\Omega_{\rm B}^2 + V_{xx} + V_{yy}\right)^2 \leq V_{xx} V_{yy} - V^2_{xy}.
\label{eq_cor_stability}
\end{equation}

In the classical case of a mass anomaly with $q=1$, 
this implies that the Lagrange points $L_4$ and $L_5$ 
are linearly stable if the Gascheau-Routh criterion $\mu \leq 0.0385...$ is met. In our cases,
$q \leq 1$ (Table~\ref{tab_param_cha_hau_qua}), so that even larger values of $\mu$
are required for $L_4$ and $L_5$ to be become unstable.
From Eq.~\ref{eq_height_mountain}, this would correspond for instance in Chariklo's case
to mountains of unrealistic heights $h > 50$~km.
Thus, mass anomalies are not expected to create unstable corotation points $L_4$ and $L_5$.

In the ellipsoid case and using Eqs.~\ref{eq_rotating_pot}, \ref{eq_f_theta_ell} and \ref{eq_cor_stability}, 
the points $C_2$ and $C_4$ are stable as long as
\begin{equation}
q^{2/3} C_{22} \lesssim  0.006.
\label{eq_C22_critic}
\end{equation}

\end{appendix}

\end{document}